\documentclass[12pt]{article}
\usepackage{epsfig}

\begin{document}

\begin{titlepage}
\begin{flushright}
arXiv:hep-ph/yymmnn \\ UFIFT-HEP-02-24 \\ CRETE-02-13
\end{flushright}

\vspace{1.5cm}

\begin{center}
\bf{POST--INFLATIONARY DYNAMICS}
\end{center}

\vspace{0.5cm}

\begin{center}
N. C. Tsamis$^{\dagger}$
\end{center}
\begin{center}
\it{Department of Physics, University of Crete \\ 
GR-710 03 Heraklion, HELLAS.}
\end{center}

\vspace{0.3cm}

\begin{center}
R. P. Woodard$^{\ddagger}$
\end{center}
\begin{center}
\it{Department of Physics, University of Florida \\ 
Gainesville, FL 32611, UNITED STATES.}
\end{center}

\vspace{0.7cm}

\begin{center}
ABSTRACT
\end{center}
\hspace*{.3cm} 
We argue that  $\Lambda$-driven inflation must overshoot into an 
era of deflation. The deflationary period ends quickly with the 
creation of a hot dense thermal barrier to the forward propagation 
of quantum correlations from the period of inflationary particle
production. Subsequent evolution is controlled by the balance 
between the persistence of this barrier and the growth in the 
4-volume from which such correlations can be seen. This balance
can lead to power law expansion.

\vspace{1cm}

\begin{flushleft}
\end{flushleft}

\vspace{0.5cm}

\begin{flushleft}
$^{\dagger}$ e-mail: tsamis@physics.uoc.gr \\
$^{\ddagger}$ e-mail: woodard@phys.ufl.edu
\end{flushleft}

\end{titlepage}

\section{Introduction}

Although it is not obvious how to account for the recent supernovae 
data \cite{supernova}, it is by now quite clear that an adequate 
period of approximately exponential expansion -- inflation 
\cite{inflation} -- provides a simple and natural explanation 
for the homogeneity and isotropy of the large-scale observable 
universe \cite{cobe}. The explicit realization of this inflationary 
phase is usually done by a scalar field but this is not necessary. 

Consider the effective four-dimensional gravitational theory that 
emerges from the full -- and yet unknown -- quantum gravitational 
theory when we restrict physical processes to scales well below 
the Planck scale. It is governed, among other things, by general 
coordinate invariance and can be written as a series of local 
terms of increasing dimensionality in the curvature. Of these 
terms, only the lowest dimensionality ones are relevant in the
range of scales of interest: the cosmological constant and the 
Ricci scalar. On the quantum level this effective theory is not 
renormalizable but, nonetheless, is BPHZ renormalizable: 
\begin{equation}
{\cal L_{\rm GR}} = {1 \over 16 \pi G} 
\Bigl(- 2 \Lambda + R \; \Bigr) \; \sqrt{-g} 
\; + \; {\rm (counterterms)} \; . \label{Lgr}
\end{equation}
Each of the BPHZ counterterms contains an infinite and a finite 
part. The infinite parts are fixed by having to absorb the 
ultraviolet divergences that are generated order by order. Of 
the finite parts, only the lowest dimensionality ones are known 
and are fixed from the measured values of the expansion rate 
for $\Lambda$ and the Newtonian force for $G$. All the remaining 
finite parts are unknown and can only be determined from the full 
theory. However, since cosmology is determined by the infrared 
sector of the theory -- where only the lowest dimensionality 
finite parts dominate -- it is insensitive to these unknown parts. 

In the above gravitational effective theory and if $\Lambda$ is 
assumed to be positive, the ``no-hair'' theorems imply that -- 
classically -- the local geometry approaches the maximally 
symmetric solution at late times \cite{deser}. This solution 
is de Sitter spacetime and, thus, $\Lambda$-driven inflation 
is intrinsic to (\ref{Lgr}) and commences naturally in a way 
that scalar-driven inflation cannot. As long as the matter
stress-energy is finite and obeys the weak energy condition,
pre-inflationary expansion redshifts the initial matter 
stress-energy until it is dominated by the cosmological
constant $\Lambda > 0$. By contrast, scalar-driven inflation
is triggered by a random field fluctuation which must be
homogeneous over more than a Hubble volume \cite{vach1}. 
This condition is so unlikely that it has not happened even
once in the observed history of the universe.

What is hard to realize is how $\Lambda$-driven inflation can 
ever stop. In Section 2, we review perturbative results from 
the inflationary regime \cite{nctrpw1}. The main conclusion 
is that long-range correlations from inflationary infrared 
graviton production build-up and screen the observed expansion 
rate. While it is the graviton that is involved in the physical 
case, we shall use -- solely for reasons of quantitative 
simplicity -- a model based on a self-interacting massless 
minimally coupled scalar which is fine-tuned to mimic most 
of the essential properties of the graviton. 

In Section 3, we argue that this screening effect inevitably 
results in a period of deflation. Stability would forbid any 
recovery on the classical level. However, the quantum universe 
is driven back to expansion because deflation compresses the 
most recently produced infrared particles during inflation into 
a hot, dense ``thermal barrier'' . This barrier degrades 
correlations from the inflationary period by scattering the 
infrared quanta that carry these correlations and, hence,
depleting the number of such quanta. 

Section 4, addresses the post-deflationary evolution of the universe.
Under the assumption that the barrier actually becomes thermal, it 
is shown that the asymptotic geometry can be that of a small power
law expanding universe. Our conclusions comprise Section 5. \\

\section{The Inflationary Regime}

{\it (i) Basics} \\
The effective theory defined by (\ref{Lgr}) is studied in the
presence of homogeneous and isotropic backgrounds:
\begin{eqnarray}
ds^2 & = & 
- dt^2 + a^2(t) \; d{\vec x} \cdot d{\vec x} 
\; = \; 
- dt^2 + 
\exp \left[ \, 2 \, b(t) \, \right]
\; d{\vec x} \cdot d{\vec x} 
\;\; , \label{ds2a} \\
& = & 
{\it \Omega}^2(\eta) \;
\Bigl( -d\eta^2 + d{\vec x} \cdot d{\vec x} \Bigr) 
\;\; , \label{ds2b}
\end{eqnarray}
where we have expressed the line element both in co-moving and
conformal coordinates. The theory is valid as long as we restrict
physics to scales below the Planck mass or, equivalently, as long 
as its dimensionless coupling constant is small:
\begin{equation}
M \equiv \Bigl( {\Lambda \over {8 \pi G}} \Bigr)^{\frac14} < M_{\rm Pl} 
\quad \iff \quad G\Lambda < 1 
\;\; . \label{validity}
\end{equation}
This is a quite wide range of scales; for instance, if $M \sim 10^{16}
GeV$ we get that $G \Lambda \sim 10^{-12}$.

Let $\gamma_{\mu\nu}(t, {\vec x})$ stand for the metric operator.
We shall be interested in quantum corrections to the classical
background geometry:
\begin{equation}
\left\langle \, state \, \left\vert \; 
\gamma_{\mu\nu}(t,{\vec x}) \; dx^{\mu} dx^{\nu} 
\; \right\vert \, state \, \right\rangle 
\; = \;
g_{\mu\nu}(t,{\vec x}) \; dx^{\mu} dx^{\nu} 
\;\; . \label{Qgmn}
\end{equation}
Geometrically significant differences between the classical and 
quantum backgrounds can be ascribed to a quantum-induced stress 
tensor. This is defined from the deficit by which the quantum 
background $g_{\mu\nu}$ fails to obey the classical equations 
of motion:
\begin{equation}
8 \pi G \; T_{\mu\nu} \equiv 
R_{\mu\nu} - \frac12 g_{\mu\nu} R + g_{\mu\nu} \Lambda 
\;\; . \label{defTmn}
\end{equation}
In (\ref{defTmn}) $ R_{\mu\nu}$ and $R$ are the Ricci tensor and
Ricci scalar constructed from the quantum background $g_{\mu\nu}$.
The structure of the induced stress tensor is dictated by homogeneity
and isotropy:
\begin{equation}
T_{00}(t) \ = \ - \rho(t) \; g_{00} 
\quad , \quad T_{0i}(t) \ = \ 0 
\quad , \quad T_{ij}(t) \ = \ p(t) \; g_{ij} 
\;\; , \label{defrho,p} 
\end{equation}
where $\rho(t)$ is the induced energy density and $p(t)$ the induced 
pressure. The latter obey:
\footnote{A dot indicates differentiation with respect to co-moving 
time $t$ while a prime denotes differentiation with respect to 
conformal time $\eta$.}
\begin{eqnarray}
\rho(t) & = & 
{1 \over 8 \pi G} \left[ \,
\dot{b}^2(t) - \Lambda \, \right]  
\;\; , \label{nonpertrho} \\
p(t)    & = & 
{1 \over 8 \pi G} \left[ \,
-2 \ddot{b}(t) \, \right] - \rho(t)
\;\; . \label{nonpertp} 
\end{eqnarray}
An observable which measures the expansion rate of the universe is the 
effective Hubble parameter: 
\begin{equation}
H(t) \equiv {{\dot a}(t) \over a(t)} = {\dot b}(t) = 
\frac{{\it \Omega}'(\eta)}{{\it \Omega}^2(\eta)} =
\sqrt {\, \frac{\Lambda}3 + \frac{8 \pi G}3 \rho(t)}
\;\; . \label{defH}
\end{equation}

{\it (ii) Results During Inflation} \\
According to \cite{deser}, time evolution from generic initial value
data leads to a locally de Sitter background: 
\begin{equation}
a_{dS}(t) = 
\exp \left[ \, b_{dS}(t) \, \right] 
= e^{Ht} = \,
{\it \Omega}_{dS}(\eta) = 
- \frac{1}{H\eta} 
\;\; , \label{dSdef}
\end{equation}
where the Hubble constant $H$ is defined as:
\begin{equation}
H^2 \equiv \frac13 \Lambda > 0
\;\; . \label{hubcondef}
\end{equation}
It is quite unlikely for inflation to have started simultaneously 
over a region larger than a causal volume. Hence, we shall consider
(\ref{dSdef}) on the manifold $T^3 \times \Re$, with the physical 
distances of the toroidal radii equal to a Hubble length at the 
onset of inflation:
\begin{equation}
- \frac12 H^{-1} < x^i \leq \frac12 H^{-1}
\;\; . \label{T3}
\end{equation}
The onset of inflation is taken to be at $t=0$:
\begin{equation}
b_{dS}(0) = 0 \quad , \quad 
{\dot b}_{dS}(0) = H
\;\; , \label{ivp1}
\end{equation}
\begin{equation}
\Big\vert \, state \Big\rangle = 
\Big\vert \; {\rm Bunch \; Davies \; vacuum \; at} \; t=0 \; 
\Big\rangle \equiv 
\Big\vert \hspace{0.05cm} 0 \Big\rangle 
\;\; . \label{ivp2}
\end{equation}
Time evolution is determined from the dynamics of the theory.
Concrete results have been obtained in a perturbation theory 
organized in terms of the fluctuating field $\psi_{\mu\nu}(x)$:
\begin{equation}
\gamma_{\mu \nu} \equiv \; 
{\it \Omega}^2 
\left( \eta_{\mu \nu} + \kappa \, \psi_{\mu \nu} \right) 
\quad , \quad 
\kappa^2 \equiv 16 \pi G
\; \; , \label{psimn}
\end{equation}
and in the infrared limit of a large number of inflationary 
e-foldings. For the background (\ref{dSdef}), the map between 
the relevant co-moving and conformal time intervals obeys:
\begin{equation}
t \in \left[ \; 0 \; , \; +\infty \right)
\quad \iff \quad
\eta \in \left[ -H^{-1} \; , \; 0^{-} \right)
\;\; . \label{timesmap}
\end{equation}

In terms of the dimensionless coupling constant $\varepsilon$ of
the theory:
\begin{equation}
\varepsilon \equiv {G \Lambda \over 3\pi} = 
\left( {\kappa H \over 4 \pi} \right)^2 = 
{8 \over 3} \left( {M \over M_{\rm Pl}} \right)^4 
\;\; , \label{epsilondef}
\end{equation}
the leading infrared results to the first non-trivial order 
of perturbation theory are \cite{nctrpw1}: 
\footnote{The higher order estimates can be found in 
\cite{nctrpw2}.}
\begin{eqnarray}
\rho_{dS}(t) & = & - \varepsilon H^4 
\left[ \; \frac{1}{8\pi^2} \; (H t)^2 \; + \; O(Ht) \; \right] 
+ \; O(\varepsilon^2) 
\;\; , \label{pertrho} \\
p_{dS}(t) & = & \varepsilon H^4 
\left[ \; \frac{1}{8\pi^2} \; (H t)^2 \; + \; O(Ht) \; \right] 
+ \; O(\varepsilon^2) 
\;\; , \label{pertp} \\
H_{dS}(t) & = & H \left\{ 1 - \varepsilon^2 
\left[ \; \frac16 \; (H t)^2 \; + \; O(Ht) \; \right] 
+ \; O(\varepsilon^3) \right\} 
\;\; . \label{pertH} 
\end{eqnarray}
The negative sign in (\ref{pertrho}) and (\ref{pertH}) is due to 
the universal attractive nature of the gravitational interaction. 
The fact that the induced energy density and pressure obey the
equation of state of vacuum energy (\ref{pertrho}-\ref{pertp}) 
is a simple consequence of stress-energy conservation:
\begin{equation}
{\dot \rho_{dS}} = - 3H \left( \, \rho_{dS} + p_{dS} \right)
\;\; , \label{tmnconservation}
\end{equation}
and the inherent weakness of the gravitational interaction for
scales $M$ below the Planck mass. This implies that $\rho_{dS}$
changes much slower than the expansion rate, and hence:
\begin{equation}
\vert {\dot \rho_{dS}}(t) \vert \; \ll 
\; H \; \vert \rho_{dS}(t) \vert
\quad \Longrightarrow \quad
p_{dS}(t) \; \sim \; - \rho_{dS}(t)
\;\; . \label{eqnstate}
\end{equation}

It becomes apparent from (\ref{pertH}) that the rate of expansion 
decreases by an amount which becomes non-perturbatively large at 
late times. This occurs when the effective coupling constant 
becomes of order one and gives a rough estimate for the number 
of inflationary e-foldings: 
\begin{equation}
\varepsilon \; Ht_1 \sim 1
\quad \Longrightarrow \quad
H t_1 \sim \left( {M_{\rm Pl} \over M} \right)^4 
\;\; . \label{N2cr}
\end{equation}
The higher order perturbative effects make a phenomenologically
insignificant correction of the rough estimate for $N_1$ 
\cite{nctrpw2}:
\begin{equation}
N_1 \equiv H t_1 \sim 
\left( {M_{\rm Pl} \over M} \right)^{\frac83}
\quad \Longrightarrow \quad N_1 \gg 60
\;\; . \label{Ncr}
\end{equation}
For any acceptable scale $M$, (\ref{Ncr}) trivially satisfies  
the bound imposed by the causality problem that inflationary 
cosmology must solve: $N_1 \gg 60$. 
\footnote{If $M \sim 10^{16} GeV$, we obtain $N_1 \sim 10^8$
which is comfortably above 60. Lower scales inflate even longer.}
Thus, $\Lambda$-driven inflation persists over many e-foldings 
for the simple reason that gravity is a weak interaction.

Moreover, the perturbative results can be used to show that 
inflation ends suddenly over the course of very few e-foldings 
\cite{nctrpw2} and, therefore, subsequent reheating becomes 
possible. Finally, the same results allow the computation of the 
spectrum of cosmological density perturbations \cite{nctrpw3} 
which are observational remnants from this epoch.

The very existence of a bare positive cosmological constant in 
the effective gravitational theory is enough to drive inflation 
-- there is no need to introduce a scalar field and to fine tune
$\Lambda$ to zero. The resulting inflationary model depends on 
the single parameter $G\Lambda$ only and, consequently, it is 
very predictive and natural. Both of these properties are severely 
diminished in scalar-driven inflation by the presence of the 
parameter space of the scalar potential.

The duration of inflation is naturally long since the gravitational 
processes must take a while to coherently superpose and overcome 
their inherently weak coupling constant $G\Lambda$. This should be 
contrasted with scalar-driven inflation, where the scalar potential 
has to be tuned to be constant over the very extended region that 
must cover at least 60 inflationary e-foldings. In our case, the 
bare cosmological constant naturally achieves the same goal since 
it {\it is} a constant.

Similarly, a sudden exit from inflation implies that the scalar 
potential must be further tuned to have a steep drop after its 
extended constancy. For $\Lambda$-driven inflation, as we approach 
the critical number of e-foldings $N_1$, the quantum gravitational 
effect gets strong and quickly screens the bare cosmological 
constant naturally. 

Another requirement for the occurence of scalar-driven inflation,
is the presence of a homogeneous fluctuation of the scalar field 
over more than one Hubble volume \cite{vach1}. This is very 
improbable and the issue is altogether avoided when gravity is 
the driving force. \\

{\it (iii) Matter Contributions During Inflation} \\
In addition to the pure gravitational sector (\ref{Lgr}), the 
effective theory below $M_{\rm Pl}$ contains a matter sector
${\cal L_{\rm M}}$:
\begin{equation}
{\cal L}_{\rm TOT} = {\cal L_{\rm GR}} + {\cal L_{\rm M}}
\;\; . \label{Ltot}
\end{equation}
All matter quanta reside in ${\cal L_{\rm M}}$ and we must 
study their back-reaction during the inflationary period, 
just as we did for the graviton. We need only be concerned 
with effectively massless particles since it is on the infrared 
limit of late observational times that we focus. This is because 
infrared effects influence a local observation through the 
coherent superposition of distant interactions in the past 
lightcone of the observer. These interactions cannot be 
transmitted by massive quanta as their propagators oscillate 
inside the lightcone and, thus, result in destructive interference.
             
The critical mass scale $m_1$ which separates an effectively 
massless from an effectively massive particle reflects whether 
the particle induces a strong infrared effect faster {\it or} 
slower respectively with respect to the graviton. An accurate 
estimate can be obtained by computing the particle mass below 
{\it or} above which the free mode functions of the particle 
suffer small {\it or} large distortions respectively over the 
relaxation time $t_1$ \cite{nctrpw4}:
\begin{equation}
m_1 \sim 
M \left( {M_{\rm Pl} \over M} \right)^{\frac73}
\;\; . \label{Mcr} 
\end{equation}

Let us now consider massless quanta. If the interactions of a
massless particle have local conformal invariance, then the 
particle is unaware of the inflating spacetime and cannot 
induce a back-reaction. This is most transparent in conformal 
coordinates where any local quantity has the same form as in 
flat space; the overall effect is nill since the conformal 
coordinate volume is only $H^{-4}$. Consequently, all effectively 
massless fermions and gauge bosons need not be considered. 
\footnote{In any case, fermions could never lead to such a 
secular effect because they obey Fermi statistics.}
The same holds true for conformally coupled effectively
massless scalars.

Effectively massless, minimally coupled scalar particles 
-- with or without self-interactions -- require closer 
inspection \cite{starob1}. Consider such a scalar theory 
in the presence of the inflationary background (\ref{dSdef}):
\begin{equation}
{\cal L}_{\varphi} =
- \frac12 \sqrt{-g} \; g^{\mu\nu} \; 
\partial_{\mu} \varphi \; \partial_{\nu} \varphi
\; - \; \sqrt{-g} \; V(\varphi)
\;\; , \label{Lmms}
\end{equation}
where, for stability reasons, the potential $V(\varphi)$ 
is to be bounded from below. Let us assume that at $t=0$
the scalar field is at the true minimum of the potential.
This minimum need not be identical to the classical minimum 
as it includes potentially non-trivial ultraviolet corrections. 
Evolution generates a finite time-dependent part in the 
vacuum expectation value of the square of the field operator: 
\footnote{This result was first obtained in \cite{phi2x}.}
\begin{equation}
\left\langle 0 \left\vert \; \varphi^2(x) \;
\right\vert 0 \right\rangle 
\; \sim \; H^2 (Ht) \; + \; 
{\rm (constant \; ultraviolet \; part)}
\;\; . \label{phi2x}
\end{equation}
Consequently, the energy of the scalar field increases with
time and so does the expansion rate.

As soon as the scalar field moves upwards its potential, the 
tendency to reverse its motion and lower its energy appears; 
it results in a decrease of the expansion rate. The two opposing 
tendencies eventually balance at a constant asymptotic value 
which has the true minimum as its lower bound. The net effect 
on the expansion rate is that it cannot decrease from its 
initial value $H$. 
\footnote{An explicit example of the generic argument is the
purely quartic potential \cite{starob1}.} 

In any case, the physical existence of scalar fields with 
dynamics based on (\ref{Lmms}) is hard to justify unless
they have a mass of the same order as the scale $M$ of the 
theory. The purpose of the above argument was to show that, 
even if they exist and are effectively massless, they cannot 
diminish the expansion rate of the universe. \\

\begin{figure}
\centerline{\epsfig{file=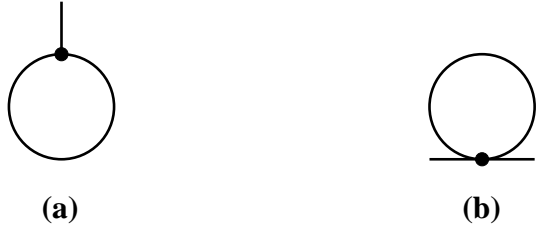,height=1.2in}}
\caption{\footnotesize First order graviton contributions to: 
{\bf (a)} the expansion rate $H(t)$ , {\bf (b)} its
\break \mbox{} \hspace{1.8cm} 
mass.}
\end{figure}

{\it (iv) Graviton-Like Scalar Model} \\
By virtue of its masslessness and lack of conformally
invariant interactions, the graviton is the unique known 
particle which can mediate a non-trivial back-reaction
to an inflationary expansion. To follow the evolution
beyond the inflationary regime, it is very instructive
-- before using the exceedingly complicated graviton
dynamics -- to construct a scalar field theory that
simulates the behaviour of the graviton. Since a free
graviton has the same dynamics as a massless minimally 
coupled scalar, it is reasonable to use the latter with 
a suitable interaction. Although a derivative interaction 
would perhaps be more appropriate, we shall choose a quartic 
self-interaction for calculational simplicity:
\begin{equation}
{\cal L}_{\varphi} =
- \frac12 \sqrt{-g} \; g^{\mu\nu} \; 
\partial_{\mu} \varphi \; \partial_{\nu} \varphi \; - \; 
\frac{1}{4!} \sqrt{-g} \; \lambda \varphi^4
\; + \; {\rm (counterterms)}
\;\; . \label{Lphi4}
\end{equation}
There are two basic properties of the graviton we should 
like the scalar theory to retain to the extent that it can: \\
-- The scalar should induce an infrared effect which reduces 
   the inflationary expansion of spacetime. \\
-- The scalar should stay massless as much as possible. 

Neither of these properties are present in (\ref{Lphi4}). 
The source of the difference resides in the part of the coincidence 
limit of the propagator that grows linearly with time: in contrast 
to the scalar, this part does not couple to anything at one loop 
for the graviton. Hence, there is no first order contribution to 
$H(t)$ from gravitons \cite{ford1}. Gravitational infrared 
contributions to $H(t)$ begin at second order and are a net drag 
on the expansion rate of the universe \cite{nctrpw2}. No graviton 
mass develops at any order. {\it (see Figure 1)}

In the scalar field theory (\ref{Lphi4}), there is a first 
order ultraviolet contribution to $H(t)$ that enhances the 
expansion of spacetime. Furthermore, to the same order, there 
is a time dependent scalar mass induced; the mass counterterm 
can only absorb the constant ultraviolet part of (\ref{phi2x}).
{\it (see Figure 2)}

\begin{figure}
\centerline{\epsfig{file=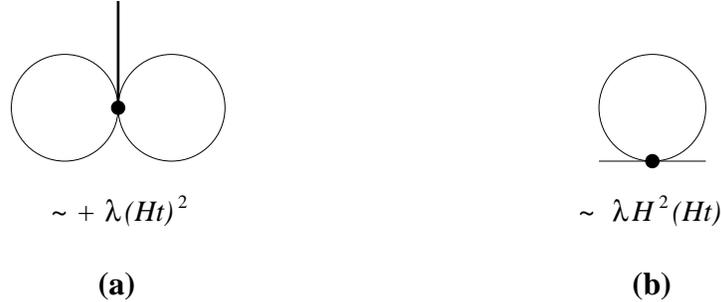,height=1.6in}}
\caption{\footnotesize First order scalar contributions to: 
{\bf (a)} the expansion rate $H(t)$ , {\bf (b)} its
\break \mbox{} \hspace{1.8cm} 
mass.}
\end{figure}

To achieve the desired correspondence, we must normal order
the scalar theory. More precisely, since we operate in curved 
spacetime, we must {\it covariantly} normal order:
\begin{equation}
\colon \varphi^2(x) \; \colon \; \equiv \;
\varphi^2(x) - i\Delta_g (x;x) 
\;\; , \label{cnophi2}
\end{equation}
where the subscript $g$ reminds us of the dependence of $\Delta(x;x)$
on the metric $g_{\mu\nu}$. It follows that:
\begin{eqnarray}
{\cal L}_{\colon \varphi \colon} 
& \equiv &
- \ \frac12 \ \sqrt{-g} \; g^{\mu\nu} \; 
\colon \partial_{\mu} \varphi \; \partial_{\nu} \varphi \; \colon
\; - \; \frac{1}{4!} \sqrt{-g} \; \lambda \; \colon \varphi^4 \; \colon
\nonumber \\
& \mbox{} &    
+ \;\; {\rm (counterterms)}
\;\; , \\
& = &
- \ \frac12 \ \sqrt{-g} \; g^{\mu\nu} \; 
\left[ \; \partial_{\mu} \varphi \; \partial_{\nu} \varphi \; - \;  
{\overrightarrow \partial_{\mu}} \; i\Delta_g(x;x) \;
{\overleftarrow \partial_{\nu}} \; \right] 
\nonumber \\
& \mbox{} &    
- \ \frac{1}{4!} \ \sqrt{-g} \; \lambda \; 
\left[\;  \varphi^4 - 6i\Delta_g(x;x) \; \varphi^2 + 
3 \left( i\Delta_g(x;x) \right)^2 \; \right]
\nonumber \\
& \mbox{} &    
+ \;\; {\rm (counterterms)}
\;\; . \label{cnoLphi4}
\end{eqnarray}
In the flat spacetime limit, the above prescription reduces 
to the usual normal ordering one.

The scalar theory given by (\ref{cnoLphi4}) is fully well-defined
when the metric is non-dynamical. For a dynamical metric, it is
only well-defined perturbatively \cite{rpw1}.
\footnote{Non-perturbatively, (\ref{cnoLphi4}) is ill-defined 
when taken as a fundamental theory with a dynamical graviton.
Then, the covariant normal ordering prescription introduces 
non-locality since the propagator is a non-local functional
of the metric and is part of the Lagrangian.}
In (\ref{cnoLphi4}) there are no more scalar tadpoles. Their
elimination automatically implies that there can be no first 
order expansion enhancing contribution to $H(t)$, and no first 
order time dependent scalar mass generated. 
\footnote{However, unlike the graviton which is protected by 
general coordinate invariance, such a scalar mass will be 
generated at second order.} 
Moreover, the conservation of the scalar stress tensor is 
preserved because of the covariant nature of the normal ordering 
prescription.  

\begin{figure}
\centerline{\epsfig{file=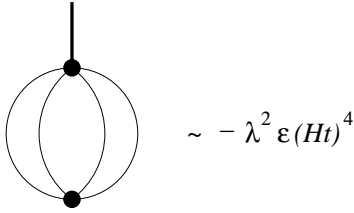,height=1.1in}}
\caption{\footnotesize Second order scalar contributions to the 
expansion rate $H(t)$.} 
\end{figure}

The first non-trivial contributions to $H(t)$ have been 
computed \cite{nctrpw4} and they behave in the desired way 
by slowing the inflationary expansion {\it (see Figure 3)}. 
In fact, for reasonable values of $\lambda$, they do so by an 
amount which overwhelms the analogous reduction (\ref{pertH}) 
due to the graviton: 
\begin{equation}
H^{\colon \varphi \colon}_{dS}(t) = H \left\{ 1 - 
\frac{\lambda^2 \, \varepsilon}{2^7 \, \pi^4}
\left[ \; \frac{1}{3^3} \; 
(H t)^4 \; + \; O(H^3t^3 ) \; \right] 
+ \; O(\lambda^3) \; \right\} 
\;\; . \label{pertHcnoscalar}
\end{equation}
This result should be contrasted with that coming from the 
scalar field theory (\ref{Lphi4}) which is not covariantly
normal ordered \cite{rpw2}:
\begin{equation}
H^{\varphi}_{dS}(t) = H \left\{ 1 +
\frac{\lambda \, \varepsilon}{2^4 \, \pi^2}
\left[ \; \frac13 \; 
(H t)^2 \; + \; O(Ht) \; \right] 
+ \; O(\lambda^2) \; \right\} 
\;\; . \label{pertHscalar}
\end{equation}

\section{Deflation Begins}

{\it (i) General Analysis} \\
Inflation is slowed by the quantum gravitational response to 
the production of infrared gravitons. However, to understand 
post-inflationary evolution we must develop a manner of thinking 
about the various physical effects which can be applied for a
general scale factor. 

Since physical gravitons obey the same equation of motion as 
massless, minimally coupled scalars, we shall study the 
Lagrangian density:
\begin{equation}
{\cal L} =
- \frac12 \sqrt{-g} \; g^{\mu\nu} \; 
\partial_{\mu} \varphi \; \partial_{\nu} \varphi 
\;\; , \label{Ldens}
\end{equation}
on the manifold $T^3 \times \Re$ with spatial co-moving 
coordinate range (\ref{T3}). Wave vectors are discrete:
\begin{equation}
{\vec k} = 2\pi H {\vec n}
\;\; , \label{discretek}
\end{equation}
and phase space integrals become mode sums:
\begin{equation}
\int \frac{d^3k}{(2\pi)^3} \; f(\vec k) 
\quad \longrightarrow \quad
H^3 \; \sum_{\vec k} f(\vec k) =
H^3 \; \sum_{\vec n} f(2\pi H {\vec n})
\;\; . \label{modesums}
\end{equation}
The scalar field can be expanded as a spatial Fourier sum:
\begin{equation}
\varphi(\eta, {\vec x}) \; = \; H^3 
\; \sum_{\vec k} \, 
e^{i {\vec k} \cdot {\vec x}} \,\,
{\widetilde \varphi}(\eta, {\vec k})
\;\; , \label{ftphi} 
\end{equation}
and each wave number can be treated separately because the 
Lagrangian is diagonal in the mode basis:
\begin{eqnarray}
L(\eta) 
& \equiv & 
\int d^3x \;\; {\cal L}(\eta, {\vec x}) 
\label{Ldef} \\
& = & 
\frac12 \, H^3 \, {\it \Omega}^2 \; 
\sum_{\vec k} \;
\left[ \, 
\Vert \, {\widetilde \varphi}'(\eta, {\vec k}) \, \Vert^2  \; - \;
k^2 \, \Vert \, {\widetilde \varphi}(\eta, {\vec k}) \, \Vert^2 
\, \right] 
\; \; . \label{Lphi}
\end{eqnarray}
The momentum canonically conjugate to 
${\widetilde \varphi}(\eta, {\vec k})$ is:
\begin{equation}
\pi(\eta,\vec{k}) \, \equiv \
H^3 \, {\it \Omega}^2 \; 
{\widetilde \varphi}^{\hspace{0.03cm} *}{'}(\eta, {\vec k}) 
\;\; , \label{momentum}
\end{equation}
resulting in the following generator of conformal time evolution 
for wave number ${\vec k}$ :
\begin{eqnarray}
H(\eta,{\vec k}) 
& = & 
{\Vert \, \pi(\eta, {\vec k}) \, \Vert^2 
\over 2 H^3 \, {\it \Omega}^2} 
\, + \, 
\frac12 \, H^3 \, {\it \Omega}^2 \, k^2 \,
\Vert \, {\widetilde \varphi}(\eta, {\vec k}) \, \Vert^2
\label{Hcanonical} \\
& = & 
\frac12 \, H^3 \, {\it \Omega}^2 
\left[ \, 
\Vert \, {\widetilde \varphi}'(\eta, {\vec k}) \, \Vert^2  
\; + \;
k^2 \, \Vert \, {\widetilde \varphi}(\eta, {\vec k}) \, \Vert^2 
\, \right] 
\label{Hconfiguration}
\end{eqnarray}

At any instant ${\it \Omega}(\eta)$ is just a number. By comparing 
(\ref{Hcanonical}) with the usual harmonic oscillator of mass 
$m = H^3 \, {\it \Omega}^2$ and frequency $\omega = k$, we conclude 
that the minimum energy state at any particular instant has conformal 
energy equal to $\frac12 k$. Because ${\it \Omega}(\eta)$ changes 
in time, the minimum energy state is different at different values 
of $\eta$ but the minimum energy is constant. 

To understand how the conformal energy evolves for the vacuum, 
we expand the scalar field:
\begin{equation}
{\widetilde \varphi}(\eta, {\vec k}) \, = \,
u(\eta, k) \; \alpha({\vec k}) \; + \;
u^*(\eta, k) \; \alpha^{\dagger}(-{\vec k}) 
\;\; , \label{ftphiexpansion} 
\end{equation}
in time independent annihilation and creation operators 
$\alpha({\vec k})$ and $\alpha^{\dagger}(-{\vec k})$ 
satisfying:
\begin{equation}
\alpha({\vec k}) \ 
\Big\vert \hspace{0.05cm} vac \Big\rangle = 0
\;\; , \label{vacdef}
\end{equation}
and canonical commutation relations:
\begin{equation}
\left[ \; \alpha({\vec k}) \; , \;
\alpha^{\dagger}({\vec k}') \; \right]
= H^{-3} \; \delta_{{\vec k} , {\vec k}'}
\;\; . \label{ccr}
\end{equation}
The mode functions $u(\eta, k)$ obey:
\begin{equation}
u'' \; + \; 2 \;
\frac{{\it \Omega}'}{{\it \Omega}} 
\; u' \; + \; k^2 u 
\; = \; 0 
\;\; , \label{modeseqn}
\end{equation}
where the normalization is fixed by comparison between 
(\ref{momentum}) and (\ref{ccr}):
\begin{equation}
u(\eta, k) \; {u^*}'(\eta, k) \; - \;
u^*(\eta, k) \; u'(\eta, k) \, = \,
i \, {\it \Omega}^{-2}(\eta)
\;\; . \label{wronskian}
\end{equation}
Substituting (\ref{ftphiexpansion}) into $H(\eta, {\vec k})$ 
and taking the vacuum expectation value gives:
\begin{equation}
\Bigl\langle vac \hspace{0.05cm} \Bigl\vert 
\ H(\eta, \vec{k}) \ 
\Bigr\vert \hspace{0.05cm} vac \Bigr\rangle 
\; = \;
\frac12 \, {\it \Omega}^2
\left[ \, \Vert \, u'(\eta, k) \, \Vert^2 \, + \,
k^2 \, \Vert \, u(\eta, k) \, \Vert^2 \, \right]
\;\; . \label{Hvev}
\end{equation}
The physical energy $E(t, {\vec k})$ is the generator of 
co-moving time and -- by a straightforward coordinate 
transformation -- equals:
\begin{equation}
E(t, {\vec k}) \; = \;
{\it \Omega}^{-1} \, H(\eta, {\vec k}) 
\; = \;
\frac12 \; a(t) \, 
\left[ \, \Vert \, u'(\eta, k) \, \Vert^2  \, + \,
k^2 \, \Vert \, u(\eta, k) \, \Vert^2 \right] 
\;\; . \label{energy}
\end{equation} 

In (\ref{energy}), the first term always represents the 
instantaneously lowest energy, the physical energy of one 
quantum. The second term can indicate particle production 
due to the inability of long wavelength virtual quanta to 
recombine by becoming trapped in the Hubble flow. If such 
production occurs, the 0-point energy of a specific mode 
becomes significantly bigger than its minimum 0-point energy. 
Because the fundamental quantum of energy for one particle 
of wave number $\vec{k}$ is comparable to $k \, a^{-1}$, 
we can think of the number $N(t, {\vec k})$ of particles 
present as:
\begin{equation}
N(t, {\vec k}) \; \sim \;
\frac12 \, k \,\, 
\Vert \; a(t) \,\, u(\eta, k) \, \Vert^2 
\;\; , \label{numberquanta}
\end{equation}
and use it to quantify the physical distinction between 
``infrared'' modes -- whose wavelength does not allow them
to recombine and, therefore, represent real particle creation
-- and ``ultraviolet'' modes -- whose wavelength allows them
to recombine and annihilate:
\begin{eqnarray}
{\it Infrared \; Modes} \quad & \Longrightarrow & \quad
N(t, {\vec k}) \; \gg \; 1
\;\; , \label{irmodesnumber} \\
{\it Ultraviolet \; Modes} \quad & \Longrightarrow & \quad
N(t, {\vec k}) \; \sim \; 0
\;\; . \label{uvmodesnumber}
\end{eqnarray}
It is important to note that, once particle creation happens, 
$N(t, {\vec k})$ does not typically decrease.

The preceding analysis of infrared particle production is 
the same for both gravitons and scalars. Of sourse, how 
newly created quanta interact is different in gravitation 
and the graviton-like scalar model. Nonetheless, two features 
are the same. First, the interaction between correlated pairs 
-- or triads -- is much stronger than between uncorrelated 
particles. In quantum gravity, this is why the effect is so 
much stronger than simply adding the homogeneous average of 
the created particle stress to the bare cosmological term. 
Later on, we shall see that degrading this correlated 
interaction plays a crucial role in achieving stability
against runaway deflation.

The second general feature is that a local observer can only 
be affected by the particle creation which lies in his past 
light cone. Many secular effects can be understood as arising 
from the coherent superposition of interactions throughout 
the invariant volume of the past light cone:
\begin{equation}
V_{\rm PLC}(\eta) \; = \;
\int_{-\frac1{H}}^{\eta} d\eta' 
\,\, {\it \Omega}^4(\eta') \;\; 
\frac{4\pi}{3} \left( \eta -\eta' \right)^3
\;\; . \label{Vplc}
\end{equation}

{\it (ii) The Physical Picture During Inflation} \\
The physical process responsible for the expansion diminishing
quantum-induced stress tensor, are the correlated interactions
among inflationary particles produced throughout the past 
lightcone of the observer. 
\footnote{As already mentioned, inflationary particle production 
by itself furnishes a much weaker stress tensor. There is no 
secular effect generated and the expansion rate suffers a 
negligible increase.}
At the onset of inflation and in the vacuum state, all modes
are virtual. As time evolves, infrared graviton pairs continuously 
ppear and -- in contradistinction to ultraviolet pairs -- cannot 
recombine to annihilate because they get pulled by the rapid 
expansion of spacetime {\it (see Figure 4)}. While these long 
wavelength gravitons get separated and become causally disconnected, 
their long-range gravitational potentials persist since a potential
exists everywhere in the forward lightcone of its source. The 
addition of the potentials of each receding infrared graviton 
pair is a secular effect and provides the negative gravitational 
interaction energy responsible for the reduction of the expansion 
rate.

\begin{figure}
\centerline{\epsfig{file=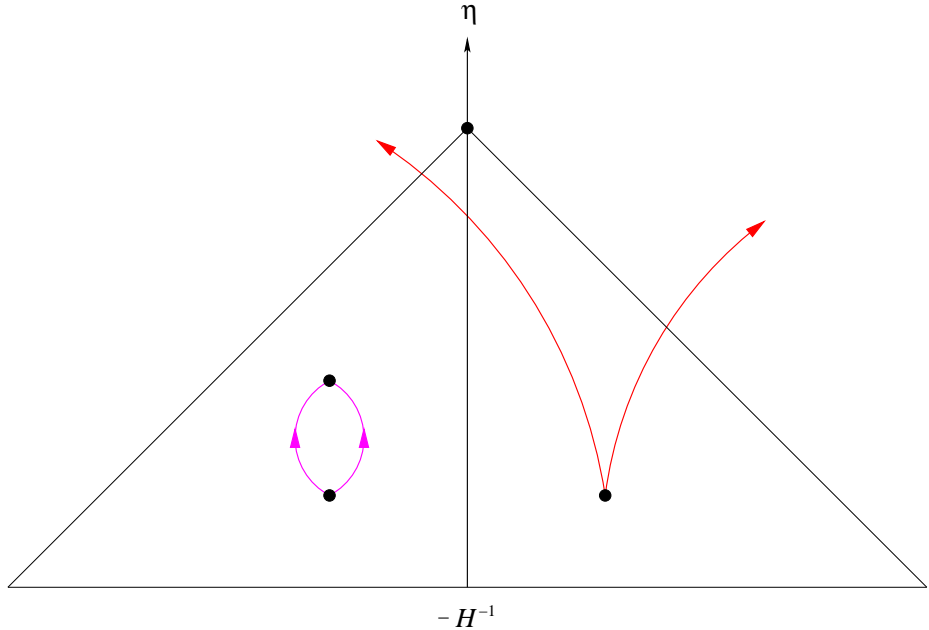,height=3.3in}}
\caption{\footnotesize Short wavelength -- $\lambda_{\rm phys} 
< H^{-1}$ -- graviton pairs {\it (violet)} recombine while
\break \mbox{} \hspace{1.8cm} 
long wavelength -- $\lambda_{\rm phys} > H^{-1}$ -- ones 
{(\it red)} cannot.}
\end{figure}

More precisely, consider the mode functions $u_{dS}(t, k)$ 
of free gravitons:
\begin{equation}
u_{dS}(t, k) \; = \;
\frac{1}{\sqrt {2k}} 
\left[ \; {\it \Omega}_{dS}^{-1}
+ \frac{iH}{k} \; \right]
e^{-ik\eta} 
\;\; . \label{inflmodefncts}
\end{equation}
It follows from (\ref{energy}) that the physical energy of 
the vacuum in mode ${\vec k}$ is:
\footnote{The 0-point energy must be consistent with the initial 
condition of an inflating background such that $H(0)=H$.}
\begin{equation}
E_{dS}(t, {\vec k}) \; = \;
\frac12 \, k \, e^{-H t} \, + \, {H^2 \, e^{H t} \over 4 k} 
\;\; . \label{inflenergy}
\end{equation}
The first term is just the -- properly redshifted -- minimum 
energy; the second term is the result of particle production. 
A typical mode begins at $t=0$ with the first term dominant. 
The second term becomes comparable at ``horizon crossing'' 
and dominates thereafter. This is the source of inflationary 
particle creation, and the onset of this enormous growth is 
what distinguishes infrared and ultraviolet modes. Horizon 
crossing of a mode occurs when its physical wave number equals 
the Hubble constant:
\begin{equation}
{\it Horizon \; Crossing} \quad \Longrightarrow \quad
k_{\rm phys} \, = \, k \; a_{dS}^{-1}(t) 
\, = \, H
\;\; , \label{horcross}
\end{equation}
and provides the physical separation between infrared
and ultraviolet modes:
\begin{eqnarray}
{\it Infrared} 
\quad & \Longrightarrow & \quad
H < k < H \, e^{Ht}
\;\; , \label{inflirmodes} \\
{\it Ultraviolet} 
\quad & \Longrightarrow & \quad
k > H \, e^{Ht}
\;\; . \label{influvmodes}
\end{eqnarray}

Furthermore, from (\ref{numberquanta}) we can obtain 
the number of gravitons created for a particular mode:
\begin{equation}
N_{dS}(t, {\vec k}) \; = \;
\frac14 
\left[ \, 1 \, + \,
\left( \frac{H \, e^{H t}}{2 k} \right)^2
\, \right] 
\;\; , \label{inflnumberquanta}
\end{equation}
and verify conditions (\ref{irmodesnumber}-\ref{uvmodesnumber}):
\begin{eqnarray}
{\it Infrared}
\quad & \Longrightarrow & \quad
N_{\rm IR}^{dS}(t, {\vec k}) \, \sim \,
\left( \frac{H \, e^{H t}}{2 k} \right)^2 
\; \gg \; 1
\;\; , \label{inflirmodesnumber} \\
{\it Ultraviolet}
\quad & \Longrightarrow & \quad
N_{\rm UV}^{dS}(t, {\vec k}) \, \sim \, 0 
\;\; . \label{influvmodesnumber}
\end{eqnarray}
Finally, it is useful to record the infrared and ultraviolet
limits of the mode functions (\ref{inflmodefncts}):
\begin{eqnarray}
{\it Infrared}
\quad & \Longrightarrow & \quad
u_{\rm IR}^{dS}(t, k) \, \sim \, 
\frac{i H}{\sqrt {2k^3}} 
\;\; , \label{inflirmodefncts} \\
{\it Ultraviolet}
\quad & \Longrightarrow & \quad
u_{\rm UV}^{dS}(t, k) \, \sim \, 
\frac{{\it \Omega}_{dS}^{-1}}{\sqrt {2k}} \; e^{-ik\eta} 
\;\; . \label{influvmodefncts}
\end{eqnarray}

To get a sense of the density of infrared gravitons present 
per causal volume, we express the number of Hubble volumes 
in terms of the initial condition $H(0)=H$:
\begin{equation}
N_H(t) \; = \; 
\left[ \; \frac{H(t) \; a(t)}{H} \; \right]^3
\;\; . \label{volumesnumber}
\end{equation}
For the inflationary geometry, we trivially get:
\begin{equation}
N^{dS}_H (t) \; = \; e^{3Ht} 
\;\; , \label{dSvolumesnumber}
\end{equation}
so that -- within one Hubble volume -- the number of any 
infrared gravitons at time $t$ is given by:
\begin{equation}
N_{\rm IR}^{dS}(t) \; = \;
\frac{1}{N^{dS}_H (t)} \times
\frac1{2\pi^2 H^3} \int_{H}^{H e^{Ht}} dk \; k^2 \times
N_{\rm IR}^{dS}(t, {\vec k}) 
\; = \; \frac{1}{8 \pi^2}
\;\; . \label{inflirparticledensity}
\end{equation}
The presence of about one infrared graviton in each Hubble 
volume implies that the initial vacuum choice and perturbation
theory is an excellent approximation during the inflationary 
regime. Such a low density cannot by itself drive a significant 
infrared screening effect; the coherent superposition from all
infrared gravitons does. \\

{\it (iii) The Inflationary Rule} \\
To make the connection of the physical picture with the quantum 
field theoretic results presented in Section 2, consider the 
first non-trivial contribution to the induced stress tensor in 
the graviton-like scalar model (\ref{cnoLphi4}) described therein. 
Its leading part is:
\footnote{Since the infrared effect derives from long wavelength 
modes, the dominant contribution to the scalar induced stress 
tensor comes from potential energy rather than kinetic energy.}
\begin{equation}
T_{\mu\nu}^{\, \colon \varphi \colon} 
\ \sim \
- \ g_{\mu\nu}^{\, dS} \; 
\left\langle 0 \left\vert 
\; \frac{1}{4!} \lambda \; \colon \varphi^4 \colon \;
\right\vert 0 \right\rangle  
\;\; . \label{leadingTmn}
\end{equation}
The relevant diagram {\it (see Figure 3)} has been calculated 
\cite{nctrpw4} in conformal coordinates where the Feynman 
rules are particularly simple. For large observation times,
the answer can be written in the following suggestive form:
\begin{equation}
T_{\mu\nu}^{\, \colon \varphi \colon} \, (\eta) 
\ \sim \;
{\it \Omega}_{dS}^2 
\; \eta_{\mu\nu} \;\;
\frac{\lambda^2 H^4}{2^9 \ 3^2 \ \pi^6} 
\; \times \; 
\ln {\it \Omega}_{dS}  
\; \times \; 
\ln^3 {\it \Omega}_{dS}  
\;\; . \label{leadingTmnresult1}
\end{equation}
In co-moving coordinates:
\begin{equation}
T_{\mu\nu}^{\, \colon \varphi \colon} \, (t) 
\ \sim \;
g_{\mu\nu}^{\, dS} \;\;
\frac{\lambda^2 H^4}{2^9 \ 3^2 \ \pi^6} 
\; \times \; (Ht) 
\; \times \; (Ht)^3
\;\; . \label{leadingTmnresult2}
\end{equation}
Expressions for the induced energy density and pressure 
follow immediately from (\ref{defrho,p}):
\begin{eqnarray}
\rho_{dS}^{\colon \varphi \colon} \, (t) & = &
- \frac{\lambda^2 H^4}{2^9 \ 3^2 \ \pi^6}  
\left[ \; (H t)^4 \; + \; O(H^3t^3 ) \; \right] 
+ \; O(\lambda^3)
\;\; , \label{leadingrho} \\
p_{dS}^{\colon \varphi \colon} \, (t) & = &
\frac{\lambda^2 H^4}{2^9 \ 3^2 \ \pi^6}  
\left[ \; (H t)^4 \; + \; O(H^3t^3 ) \; \right] 
+ \; O(\lambda^3)
\;\; . \label{leadingp} 
\end{eqnarray}

The observer has causal access to the volume of the past 
lightcone:
\begin{equation}
V_{\rm PLC}^{dS}(\eta) = 
\int_{-\frac1{H}}^{\eta} d\eta' 
\,\, {\it \Omega}_{dS}^{\, 4}(\eta') \;\; 
\frac{4\pi}{3} \left( \eta -\eta' \right)^3
\;\; , \label{inflVplc}
\end{equation}
which for large observation times becomes:
\begin{equation}
V_{\rm PLC}^{dS}(\eta) 
\; \longrightarrow \;  
\frac{4\pi}{3} H^{-4} \; \ln {\it \Omega}_{dS}  
\ = \ 
\frac{4\pi}{3} H^{-4} \; \left( Ht \right)
\;\; . \label{inflVplclimit}
\end{equation}
This is a purely geometrical quantity, independent of the 
particular dynamics of the theory. It controls the size
over which particle production can affect a local observer
and is responsible for the single $\ln {\it \Omega}_{dS} = Ht$   
factor in (\ref{leadingTmnresult1}-\ref{leadingTmnresult2}) 
since the diagram contains a single interaction vertex 
integration.

The correlations are imprinted on the propagator:
\begin{eqnarray}
i\Delta_{dS}(x; x') & = &
\frac{H^2}{8 \pi^2} \left\{
\frac{2 \eta \eta'}
{{\it \Delta} x^2 - (\vert {\it \Delta}\eta \vert - i\epsilon)^2}
- \ln \left[ H^2 \left( 
{\it \Delta} x^2 - (\vert {\it \Delta}\eta \vert - i\epsilon)^2 
\right) \right] \right\}
\;\; , \nonumber \\
& \mbox{} & 
{\it \Delta}\eta \equiv \eta - \eta' 
\quad \& \quad 
{\it \Delta} x \equiv \Vert {\vec x} - {\vec x}^{\, \prime} \Vert
\;\; . \label{dSprop}
\end{eqnarray}
Particle production resides in the logarithmic term of the
propagator; its absence reduces (\ref{dSprop}) to the flat space 
situation for which there is no such production. The propagator 
also carries the long-range field that emerges from the particles 
produced. In (\ref{leadingTmnresult1}-\ref{leadingTmnresult2}) the 
propagator is responsible for the factor 
$\ln^3 {\it \Omega}_{dS} = (Ht)^3$   
since the diagram represents an expectation value and, therefore, 
its leading part contains the cube of the real part of (\ref{dSprop}).
\footnote{The formalism for computing quantum field theoretic
expectation values was developed in \cite{inin} and was used 
extensively in \cite{nctrpw1}, \cite{nctrpw4}, \cite{rpw1}, 
\cite{rpw2}.}

The induced stress tensor can be thought of as the combined 
result of a geometrical effect -- the causal volume accessed 
by the observer -- and a dynamical effect -- the interaction 
stress among the particles produced:
\begin{eqnarray}
& \mbox{} &
\hspace{-1.7cm} 
{\it Inflationary \; Induced \; Stress} \; \sim 
\label{inflTmninwords} \\
& \mbox{} &
\hspace{-1.7cm} 
{\it ``Causal \; Volume"} \; \times \;
{\it ``Interactions \; Among \; Particles \; Produced"} 
\;\; . \nonumber 
\end{eqnarray}
The effect is inherently non-local, as it should be. Otherwise, 
it could not screen $\Lambda$; it would be absorbable into a 
re-definition of $\Lambda$. \\

{\it (iv) Deflation is Unavoidable} \\
Deflation must occur because, as the inflationary expansion 
rate is reduced, the visible region of inflationary particle 
production continues to increase at an even faster pace {\it and}
correlations remain basically the same.
\footnote{Except for flat spacetime, the logarithmic part of 
the propagator (\ref{dSprop}) is present in as wide a class 
of backgrounds (\ref{ds2a}) as it is possible to compute 
analytically \cite{nctrpw5}.}
The screening effect grows stronger and makes the expansion 
rate slow even more. There simply is no barrier preventing 
the universe to enter a contracting phase. This behaviour of 
the screening mechanism is a generic gravitational instability. 
A familiar example is the instability of black holes emitting 
Hawking radiation: as they emit radiation their temperature 
increases and makes them emit even more.

We use Latin characters to distinguish deflationary quantities 
from their inflationary counterparts:
\begin{equation}
{\rm a}(t) = 
\exp \left[ \, {\rm b}(t) \, \right] = 
\Omega (\eta) 
\quad , \quad 
{\dot {\hspace{0.15cm} {\rm b}}}(t) < 0 
\;\; . \label{defldef}
\end{equation}
The spacetime geometry (\ref{defldef}) results in a completely 
unacceptable classical cosmology. The total energy density 
$\rho_{\rm T}$ and pressure $p_{\rm T}$ of {\it any} classically 
stable theory in a homogeneous and isotropic universe must obey 
the weak energy condition \cite{hawkellis}:
\begin{equation}
\rho_{\rm T} + p_{\rm T} \; \geq \; 0
\;\; . \label{wec}
\end{equation}
From the evolution equation:
\begin{equation}
- 2 {\ddot {\hspace{0.1cm} {\rm b}}} 
\; = \; 
8 \pi G \left( \rho_{\rm T} + p_{\rm T} \right)
\;\; , \label{bddoteqn}
\end{equation}
this is equivalent to 
${\ddot {\hspace{0.1cm} {\rm b}}}(t) \leq 0$. Thus, 
if deflation occurs at $t=t_1$, it will continue to occur 
thereafter:
\begin{equation}
{\dot {\hspace{0.15cm} {\rm b}}}(t_1) < 0 
\quad \Longrightarrow \quad
{\dot {\hspace{0.1cm} {\rm b}}}(t) < 0 
\quad , \quad \forall \; t > t_1
\;\; , \label{deflcollapse}
\end{equation}
and the universe will suffer total collapse.

The above analysis is based entirely on classical gravity
and can only be invalidated by quantum gravitational effects. 
We shall argue that such an effect is present and leads to 
a recovery from the deflationary regime. It is worthwhile 
mentioning at this stage that such a quantum violation of 
the weak energy condition leads to an equation of state 
which cannot be produced by any cosmology based on classical
gravity:
\begin{equation}
\rho_{\rm T} + p_{\rm T} \; < \; 0
\quad \Longleftrightarrow \quad 
w < -1
\;\; , \label{wlessminusone}
\end{equation}
where $p_{\rm T} = w \rho_{\rm T}$. An equation of state with 
$w < -1$ is a quite unique and distinctive prediction with 
observational consequences. \\

{\it (v) The Physical Picture as Deflation Commences} \\
The maximally symmetric example of a contracting universe is 
the de Sitter geometry (\ref{dSdef}) with a reflected Hubble 
constant. Therefore, to retain our quantitative power, we 
idealize the geometry as perfect de Sitter inflation with 
Hubble constant $H$, followed by perfect de Sitter deflation 
with Hubble constant $-H$. The transition is assumed to occur 
at co-moving time $t = t_1$:
\begin{eqnarray}
a_{dS}(t) & = & 
\exp \left[ \, b_{dS}(t) \, \right]
\; = \; e^{Ht} 
\quad , \quad \forall \; 0 < t < t_1
\;\; , \\
{\rm a}_{\rm dS}(t) & = & 
\exp \left[ \, {\rm b}_{\rm dS}(t) \, \right]  
\; = \; e^{-H(t-2t_1)} 
\quad , \quad \forall \; t_1 < t < +\infty
\;\; . \label{infldefldef1}
\end{eqnarray}
The equivalent statement in conformal coordinates is:
\footnote{Because they serve no further purpose, henceforth, 
we shall drop the subscripts $dS$ and ${\rm dS}$.}
\begin{eqnarray}
{\it \Omega}_{dS}(\eta) & = & 
- \frac{1}{H\eta}
\quad , \quad \forall \; -H^{-1} < \eta < \eta_1
\;\; , \\
\Omega_{\rm dS}(\eta) & = &
\frac{1}{H (\eta - 2\eta_1)}
\quad , \quad \forall \; \eta_1 < \eta < +\infty
\;\; . \label{infldefldef2}
\end{eqnarray}
Since the number of inflationary e-foldings (\ref{Ncr}) is 
very large, the co-moving transition time satisfies:
\begin{equation}
N_1 \gg 60 
\quad \Longrightarrow \quad
t_1 \gg H^{-1} 
\;\; , \label{transtimevalue}
\end{equation}
or, in the conformal language:
\begin{equation}
N_1 \gg 60 
\quad \Longrightarrow \quad
\eta_1 = - H^{-1} e^{-N_1} 
\; \rightarrow \; 0^- 
\quad \sim \quad
\vert \eta_1 \vert \ll H^{-1}
\;\; . \label{conftranstimevalue}
\end{equation}

\begin{figure}
\centerline{\epsfig{file=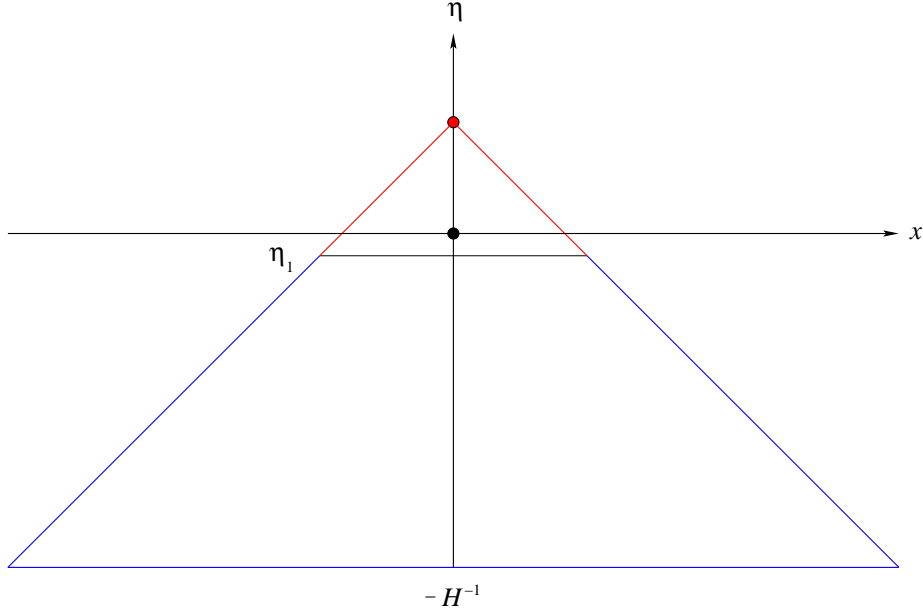,height=3.2in}}
\caption{\footnotesize The past lightcone of an observer past 
the inflation {\it (blue)} to deflation 
\break \mbox{} \hspace{1.9cm} 
{\it (red)} transition. The graph is not properly scaled.}
\end{figure}

The volume of the past lightcone of an observer at a 
post-transition time is {\it (see Figure 5)}:
\begin{eqnarray}
{\rm V}_{\rm PLC}(\eta) & = &
\int_{-\frac1{H}}^{\eta} d\eta' \; {\it \Omega}^4(\eta') \;\; 
\frac{4\pi}{3} \left( \eta -\eta' \right)^3
\;\; , \\
& = & \frac{4\pi}{3} H^{-4} \left\{ \;
\int_{-\frac1{H}}^{\eta_1} d\eta' \; 
\frac{\left( \eta -\eta' \right)^3}{\eta'^4}
+ \int_{\eta_1}^{\eta} d\eta' \; 
\frac{\left( \eta -\eta' \right)^3}
     {\left( \eta' -2\eta_1 \right)^4}
\; \right\}
\;\; , \qquad \label{vplc}
\end{eqnarray}
and for large observation times becomes:
\begin{equation}
{\rm V}_{\rm PLC}(\eta) \; \longrightarrow \;
\frac{8\pi}{9} H^{-4} \; 
\left( \frac{{\it \Omega}_1}{\Omega} \right)^3
= \ \frac{8\pi}{9} H^{-4} \; e^{3H(t-t_1)}
\;\; . \label{vplclimit}
\end{equation} 
The exponential increase with co-moving time of the past 
lightcone volume -- as opposed to the linear increase 
(\ref{inflVplclimit}) during inflation -- has an obvious
physical origin: as the universe deflates, causally 
disconnected regions come into contact again at a very 
rapid rate. Hence, some of the modes that left the 
horizon in the inflationary era, can causally interact 
again during the subsequent deflation. 

More precisely, consider the inflationary mode expansion 
of the field operator for the massless, minimally coupled 
scalar appropriate to the Bunch-Davies vacuum:
\begin{eqnarray}
{\widetilde \varphi}(\eta, {\vec k})  & = &
u(\eta, k) \; \alpha_0({\vec k}) \; + \;
u^*(\eta, k) \; \alpha_0^{\dagger}(-{\vec k}) 
\;\; , \label{inflftphiexp} \\
u(\eta, k) & = &
\frac{1}{\sqrt {2k}} 
\left[ \; {\it \Omega^{-1}}(\eta) + \frac{iH}{k} \; \right]
e^{-ik\eta}
\; \; . \label{inflmodefncts} 
\end{eqnarray}
We can retain this expansion by simply evolving the mode 
functions into the deflationary epoch. However, it is useful 
to expand the same field operator in the different basis 
appropriate to the Bunch-Davies vacuum for deflation:
\begin{eqnarray}
{\widetilde \varphi}(\eta, {\vec k})  & = &
{\rm v}(\eta, k) \; \alpha_1({\vec k}) \; + \;
{\rm v}^*(\eta, k) \; \alpha_1^{\dagger}(-{\vec k}) 
\;\; , \label{deflftphiexp} \\
{\rm v}(\eta, k) & = &
\frac{1}{\sqrt {2k}} 
\left[ \; \Omega^{-1}(\eta) - \frac{iH}{k} \; \right]
e^{-ik\eta}
\; \; . \label{deflmodefncts} 
\end{eqnarray}
The evolution of (\ref{inflmodefncts}) after the transition:
\begin{equation}
{\it Inflation} \; \longrightarrow \; {\it Deflation}
\quad \Longrightarrow \quad 
u(\eta, k) \; \longrightarrow \; {\rm u}(\eta, k)
\;\; , \label{modestransfer}
\end{equation}
is obtained by matching the mode functions and their first
derivative at $\eta = \eta_1$:
\begin{equation}
{\rm u}(\eta, k) = 
\left[ \; 1 + \frac{i H {\it \Omega}_1}{k} \; \right]
{\rm v}(\eta, k) \; + \;
\left[ \; \frac{i H {\it \Omega}_1}{k} \; 
e^{-2ik\eta_1} \; \right] 
{\rm v}^*(\eta, k)
\;\; , \label{vmodefncts}
\end{equation}
where ${\it \Omega}_1 \equiv -(H \eta_1)^{-1}$. Hence, the 
relation between the operators in the two regimes is:
\begin{equation}
\alpha_1(\vec k) = 
\left[ \; 1 + \frac{i H {\it \Omega}_1}{k} \; \right]
\alpha_0(\vec k) \; - \;
\left[ \; \frac{i H {\it \Omega}_1}{k} \; 
e^{-2ik\eta_1} \; \right] \alpha_0^{\dagger}(-{\vec k})
\;\; . \label{operatorstransfer} 
\end{equation}
The connection between the corresponding Bunch-Davies vacua:
\begin{equation}
\alpha_0 \; \Big\vert \hspace{0.05cm} 0 \Big\rangle = 0
\quad \& \quad
\alpha_1 \; \Big\vert \hspace{0.05cm} 1 \Big\rangle = 0
\;\; , \label{vacuadef}
\end{equation}
is a direct implication of (\ref{operatorstransfer}): 
\begin{equation}
\Big\vert \hspace{0.05cm} 1 \Big\rangle \sim \;
\exp{\left[ \; \frac{e^{-2ik\eta_1}}{2\;(1+ik\eta_1)} \;
{\left( \alpha_0^{\dagger} \right)^2} \; \right]} \; 
\Big\vert \hspace{0.05cm} 0 \Big\rangle 
\;\; , \label{vacuatransfer}
\end{equation}
and clearly displays the vast number of particles with
which the inflationary vacuum has been populated.

It is useful to decompose ${\rm u}(\eta, k)$ into two parts:
\begin{equation}
{\rm u}(\eta, k) \; = \;
{\rm v}(\eta, k) \, + \, {\rm U}(\eta, k)
\;\; , \label{u+Umodefncts}
\end{equation}
where (\ref{vmodefncts}) requires: 
\begin{equation}
{\rm U}(\eta, k) \; \equiv \;
\frac{i H {\it \Omega}_1}{k} \; e^{-ik\eta_1} 
\left[ \ e^{ik\eta_1} \; {\rm v}(\eta, k) +
e^{-ik\eta_1} \; {\rm v}^*(\eta, k) \ \right]
\;\; . \label{Umodefncts} 
\end{equation}
We can rewrite (\ref{Umodefncts}) as follows:
\begin{equation}
{\rm U}(\eta, k) =
{\sqrt \frac2{k}} \; \frac{H}{k} \; 
\frac{{\it \Omega}_1}{\Omega} \; ie^{-ik\eta_1} 
\left\{ 
{\cos \left[ k (\eta - \eta_1) \right]} \ - \
\frac{H \Omega}{k} \; {\sin \left[ k (\eta - \eta_1) \right]}
\right\}
\label{Umodefncts2} 
\end{equation}
Due to the infinitesimal proximity of the transition 
conformal time $\eta_1$ to $0^-$, the oscillating argument 
in (\ref{Umodefncts2}) can be excellently approximated:
\begin{equation}
k \; (\eta - \eta_1) \; \sim \; k \; (\eta - 2\eta_1) 
\; = \; \frac{k}{H \Omega} \; \equiv \; {\rm x}
\;\; , \label{xdef}
\end{equation}
so that:
\begin{equation}
{\rm U}(\eta, k) \ \sim \
{\sqrt \frac2{k}} \; \frac{H}{k} \; 
\frac{{\it \Omega}_1}{\Omega} \; ie^{-ik\eta_1} 
\left( \cos {\rm x} \ - \
\frac{\sin {\rm x}}{{\rm x}} \right)
\;\; . \label{Umodefncts3} 
\end{equation}

The number of available gravitons of a particular mode 
is obtained by substituting (\ref{vmodefncts}) in 
(\ref{numberquanta}):
\begin{eqnarray}
{\rm N}(t, {\vec k}) & = &
\frac14 \left( 1 + 2 \, \frac{H^2 {\it \Omega}_1^2}{k^2} \right)
\left( 1 + \frac{H^2 \, \Omega^2}{k^2} \right)
\nonumber \\
& \mbox{} &
+ \; \frac14 \; \frac{i H {\it \Omega}_1}{k}
\left( 1 - \frac{i H {\it \Omega}_1}{k} \right)
\left( 1 + \frac{i H \Omega}{k} \right)^2 \;
e^{2 i k (\eta - \eta_1)}
\nonumber \\
& \mbox{} &
- \; \frac14 \; \frac{i H {\it \Omega}_1}{k}
\left( 1 + \frac{i H {\it \Omega}_1}{k} \right)
\left( 1 - \frac{i H \Omega}{k} \right)^2 \;
e^{-2 i k (\eta - \eta_1)}
\;\; . \label{deflnumberquanta}
\end{eqnarray}
In a deflating geometry -- unlike an inflating one -- 
there is no horizon and, hence, no particle creation. From 
(\ref{u+Umodefncts}) and (\ref{deflnumberquanta}), which 
take into account that deflation followed an inflationary 
epoch, we can physically distinguish three kinds of modes. 
Keeping in mind that from the transition onwards 
${\it \Omega}_1 \gg \Omega$, we have: \\
$\ast$ {\it Infrared modes,} characterized by the smallest 
wave numbers and high occupancy,
\begin{eqnarray}
{\it Infrared} 
\quad & \Longrightarrow & \quad
H < k < H \, \Omega
\;\; , \label{deflirmodes} \\
\mbox{}
& \Longrightarrow & \quad
{\rm u}_{\rm IR}(t, k) \; \sim \; 
- \frac{i H}{\sqrt {2k^3}} 
\;\; , \label{deflirmodefncts} \\
\mbox{}
& \Longrightarrow & \quad
{\rm N}_{\rm IR}(t, {\vec k}) \, \sim \,
\frac{H^2 \, {\it \Omega}_1^2}{2k^2} 
\; \gg \; 1
\;\; . \label{deflirmodesnumber}
\end{eqnarray}
As (\ref{deflirmodesnumber}) indicates, infrared modes 
are not produced after $t_1$; they were created during 
inflation when they could not recombine and they continue 
to exist thereafter. A typical infrared mode 
(\ref{deflirmodes}) satisfies ${\rm x} \ll 1$ and, 
therefore, $\cos{\rm x}$ almost cancels against 
${\rm x}^{-1} \sin{\rm x}$ in (\ref{Umodefncts3}) to 
leave a negligible contribution to ${\rm U}(\eta, k)$. 
The mode functions (\ref{deflirmodefncts}) come solely 
from the ${\rm v}(\eta, k)$ part of 
(\ref{u+Umodefncts}). \\
$\ast$ {\it Ultraviolet modes,} characterized by the largest 
wave numbers and low occupancy,
\begin{eqnarray}
{\it Ultraviolet} 
\quad & \Longrightarrow & \quad
k > H {\it \Omega}_1
\;\; , \label{defluvmodes} \\
& \Longrightarrow & \quad
{\rm u}_{\rm UV}(t, k) \; \sim \; 
\frac{\Omega^{-1}}{\sqrt {2k}} \; 
e^{-ik\eta} 
\;\; , \label{defluvmodefncts} \\
& \Longrightarrow & \quad
{\rm N}_{\rm UV}(t, {\vec k}) \, \sim \, 0 
\;\; . \label{defluvmodesnumber}
\end{eqnarray}
They are virtual throughout the entire evolution. Because 
a typical ultraviolet mode (\ref{defluvmodes}) obeys 
${\rm x} \gg {\Omega}^{-1} {\it \Omega}_1 \gg 1$ , only 
the $\cos{\rm x}$ term survives in (\ref{Umodefncts2}). 
Nonetheless, it is multiplied by the ultraviolet factor 
$k^{-\frac32}$ so that the corresponding mode functions 
(\ref{defluvmodefncts}) reside exclusively in the 
${\rm v}(\eta, k)$ part of (\ref{u+Umodefncts}). \\
$\ast$ {\it Thermal modes,} characterized by the remaining 
wave numbers and -- in the typical case -- high occupancy, 
\footnote{The reason for calling these modes thermal shall 
become clear later.}
\begin{eqnarray}
{\it Thermal} 
\quad & \Longrightarrow & \quad
H \, \Omega < k < H {\it \Omega}_1
\;\; , \label{deflthermodes} \\
& \Longrightarrow & \quad
{\rm u}_{\rm TH}(t, k) \, \sim \,
{\sqrt \frac2{k}} \; \frac{H}{k} \; 
\frac{{\it \Omega}_1}{\Omega} \; 
\cos \left( \frac{k}{H \Omega} \right)
\;\; , \label{deflthermodefncts} \\
& \Longrightarrow & \quad
{\rm N}_{\rm TH}(t, {\vec k}) \, \sim \,
\frac{H^2 \, {\it \Omega}_1^2}{k^2} \;
\cos^2 \left( \frac{k}{H \Omega} \right)
\; \gg \; 1
\;\; . \qquad \label{deflthermodesnumber} 
\end{eqnarray}
These are, by virtue of their large occupation number,
``infrared'' modes during deflation. However, their
wavelength allows them to potentially recombine after 
the transition at $t_1$. From (\ref{deflthermodesnumber}), 
it is apparent that perfect annihilation occurs only when:
\begin{equation}
k_{\rm phys} \; = \; 
k \, \Omega^{-1} \, \sim \;
\pi H \left( \ell + \frac12 \, \right)
\quad , \quad 
\forall \, \ell \in {\cal N}_0
\;\; . \label{thermodesannih}
\end{equation} 
For a typical thermal mode (\ref{deflthermodes}), where 
$1 \ll {\rm x} \ll {\Omega}^{-1} {\it \Omega}_1$ , only
the $\cos{\rm x}$ term survives in (\ref{Umodefncts2}) 
and it is multiplied by ${\Omega}^{-1} {\it \Omega}_1 \gg 1$. 
Therefore, it is the thermal modes which contribute to 
${\rm U}(\eta, k)$ and that contribution is big. It is
important to note that right after the transition at $t_1$, 
${\rm u}_{\rm TH}(\eta, k)$ behaves like 
${\rm u}_{\rm IR}(\eta, k)$, that is, like a constant. 
But as soon as the contracting $\Omega$ has shrunk to 
the point that the $\Omega^{-1}$ term overwhelms the 
$i H k^{-1}$ term, ${\rm u}_{\rm TH}(\eta, k)$ starts 
to blueshift like $\Omega^{-1}$ -- in contradistinction
to ${\rm u}_{\rm IR}(\eta, k)$ which always stays constant. 
The ``crossover'' when the constant thermal mode functions
behaviour becomes a growth is given by $k \sim H \Omega$. 

Finally, a deflationary observer can measure the density 
of any thermal gravitons per invariant 3-volume:
\begin{equation}
{\rm \rho}_{\rm TH}(t)  = 
\frac{H^3}{{\rm a}^3(t)} \times
H^{-3} \int_{H \Omega}^{H {\it {\Omega}_1}} dk \; k^2 
\times {\rm N}_{\rm TH}(t, {\vec k}) \; \sim \;  
H^3 \; e^{H(t - t_1)} 
\;\; , \label{defltherdens} 
\end{equation}
and conclude that it increases exponentially with 
co-moving time. \\

{\it (vi) Perturbation Theory is Revised} \\
Perturbative computations in the presence of the state 
(\ref{ivp2}) are no longer appropriate. The density of 
thermal modes grows exponentially due to the contracting 
3-volume. A part of the quantum field is effectively 
behaving like a classical field. Thus, we can absorb 
it into the background and use the classical field 
equations -- which can be solved -- to follow its 
non-perturbative evolution.

More precisely, the full quantum field (\ref{ftphi}) is 
decomposed into two parts:
\begin{equation}
\varphi(\eta, {\vec x}) =
\phi(\eta, {\vec x}) \; + \; \Phi(\eta, {\vec x}) 
\;\; . \label{phi+Phifield}
\end{equation}
The quantum field $\phi(\eta, {\vec x})$ is the mode expansion
corresponding to a state which is in Bunch-Davies vacuum at the
start of deflation:
\begin{equation}
\phi(\eta, {\vec x}) =
\int \frac{d^3k}{(2\pi)^3} \; e^{i {\vec k} \cdot {\vec x}} \;
\left\{ \; {\rm v}(\eta, k) \; \alpha_0({\vec k}) \; + \;
{\rm v}^*(\eta, k) \; \alpha_0^{\dagger}(-{\vec k}) \; \right\}
\;\; , \label{iruvphi} 
\end{equation}
while the field $\Phi(\eta, {\vec x})$ expresses the enormous
occupancies of the thermal and infrared modes for the actual 
state:
\begin{equation}
\Phi(\eta, {\vec x}) =
\int \frac{d^3k}{(2\pi)^3} \; e^{i {\vec k} \cdot {\vec x}} \;
\left\{ \; {\rm U}(\eta, k) \; \alpha_0({\vec k}) \; + \;
{\rm U}^*(\eta, k) \; \alpha_0^{\dagger}(-{\vec k}) \; \right\}
\;\; . \label{thermalphi} 
\end{equation}
The latter commutes with its time derivative:
\footnote{This is because, unlike a usual quantum field, its 
creation and annihilation parts are both multiplied by a phase 
factor with identical time dependence.}
\begin{equation}
\left[ \; {\widetilde \Phi}(\eta, {\vec k}) \; , \,
{\widetilde \Phi}'(\eta, {\vec k}') \; \right] \; = \; 0
\;\; , \label{thermalphiccr}
\end{equation}
so that its quantum fluctuations are essentially frozen.
Strictly speaking, it is still a quantum field since it has
an expansion (\ref{thermalphi}) in terms of creation and
annihilation operators. However, by giving random assignments
to $\alpha_0$ and $\alpha_0^{\dagger}$ it can be treated,
thereafter, as a classical field. The appearance of a 
continuously increasing number of thermal modes, makes 
perturbative calculations in the presence of the state 
(\ref{ivp2}) invalid. Accordingly, the field $\Phi$ -- 
which describes the ``hot dense soup" of thermal modes -- 
should be considered as a classical background with the 
understanding that $\alpha_0$ and $\alpha_0^{\dagger}$ 
have been assigned random values. Hence, the ``hot dense 
soup'' of particles formed is taken into account by being 
treated as a classical background. It is around this 
stochastic background that perturbation theory must be 
developed. This revised perturbation theory will remain 
valid after $\Phi$ becomes strong enough for its 
self-interactions to become relevant. 

The field $\Phi$ defined by (\ref{thermalphi}) and 
(\ref{deflthermodefncts}) can be reduced as follows:
\begin{equation}
\Phi(\eta) \; \sim \;
\int_{H\Omega}^{H{\it \Omega}_1} \frac{d^3k}{(2\pi)^3} \; 
{\sqrt \frac2{k}} \; \frac{H}{k} \; 
\frac{{\it \Omega}_1}{\Omega} \; 
{\cos \left( \frac{k}{H \Omega} \right)} \; 
i \left[ \; \alpha_0({\vec k}) \; - \;
\alpha_0^{\dagger}(-{\vec k}) \; \right]
\;\; , \label{thermalphireduced} 
\end{equation}
where we have exploited the negligible spatial variation of all 
thermal modes except those with $k \sim H {\it \Omega}_1$. \\

{\it (vii) Evaluation of the Stochastic Background} \\
Since the initial state (\ref{ivp2}) is free, the associated
operators $\alpha_0$ and $\alpha_0^{\dagger}$ are stochastically
realized as independent complex Gaussian random variables on 
the manifold $T^3$ with coordinate range (\ref{T3}).
{\footnote{In our physical situation, the continuous wave vector
range we have been using throughout is an excellent approximation
of the discreteness implied by $T^3$.} 
The commutation relations (\ref{ccr}) imply that the independent 
stochastic variables $\alpha_0$ and $\alpha_0^*$ have standard 
deviation equal to $H^{-3}$. It is convenient to absorb the 
dimensions:
\begin{equation}
A_{\vec n} \; \equiv \; H^{\frac32} \; \alpha_0(\vec n)
\quad \& \quad
A^*_{\vec n} \; \equiv \; H^{\frac32} \; \alpha_0^*(\vec n)
\;\; , \label{Adef}
\end{equation}
so that the standard deviation of the rescaled variables 
is one: 
\begin{equation}
\sigma^2_{A}(\eta) \; = \; 
\sigma^2_{A^*}(\eta) \; = \; 1
\;\; . \label{Astandev}
\end{equation}

In terms of the variables (\ref{Adef}), the stochastic 
background (\ref{thermalphireduced}) equals:
\begin{equation}
\Phi(\eta) \; \sim \;
\frac{i H}{2\pi^{\frac32}} \; 
\frac{{\it \Omega}_1}{\Omega} \; 
\sum_{\vec n} \; n^{-\frac32} \;
{\cos \left( \frac{2\pi}{\Omega} n \right)} \; 
\left[ \; A_{\vec n} - A^*_{-{\vec n}} \; \right]
\;\; , \label{discretethermalphi} 
\end{equation}
where the summation range is $(2\pi)^{-1} \Omega < n <
(2\pi)^{-1} {\it \Omega}_1$ . Its standard deviation
is given by:
\begin{equation}
\sigma^2_{\Phi}(\eta) \; \sim \;
\frac{H^2}{4\pi^3} \; 
\left( \frac{{\it \Omega}_1}{\Omega} \right)^2 \; 
\sum_{\vec n} \; n^{-3} \;
{\cos^2 \left( \frac{2\pi}{\Omega} n \right)} 
\;\; , \label{thermalphistandev} 
\end{equation}
and the sum present in (\ref{thermalphistandev}) can be 
evaluated:
\begin{eqnarray}
\sum_{\vec n} \; n^{-3} \;
{\cos^2 \left( \frac{2\pi}{\Omega} n \right)} 
& \sim &  
4\pi \int_{\frac{\Omega}{2\pi}}^{\frac{{\it \Omega}_1}{2\pi}} \;
\frac{dy}{y} \; 
\frac12 \left[ \; 1 +  
{\cos \left( \frac{4\pi}{\Omega} y \right)} \; \right]
\;\; , \\
& = & 
2\pi \; \ln {\left( \frac{{\it \Omega}_1}{\Omega} \right)}
\; + \; {\it O}(1)
\;\; . \label{standevsum} 
\end{eqnarray}
Therefore, a good measure of the stochastic background 
$\Phi$ is:
\begin{equation}
\Phi(\eta) \; \sim \; 
\sqrt {\sigma^2_{\Phi}(\eta)} \; \sim \;
\frac{H}{{\sqrt 2} \, \pi} \; 
\left( \frac{{\it \Omega}_1}{\Omega} \right) \; 
\sqrt {\ln {\left( \frac{{\it \Omega}_1}{\Omega} \right)}}
\;\; . \label{thermalphivalue} 
\end{equation}

\section{The Deflationary Regime}

{\it (i) The Physical Picture During Deflation} \\
Throughout the inflationary era there was production of 
infrared gravitons with a density of about one per Hubble 
volume. As deflation proceeds, the volume contracts and 
thermal modes appear; the first to do so have wave number 
$k = H {\it \Omega}_1$. Thereafter, modes with lower wave 
numbers follow and the density of any such thermal gravitons 
per invariant 3-volume increases dramatically 
(\ref{defltherdens}).

\begin{figure}
\centerline{\epsfig{file=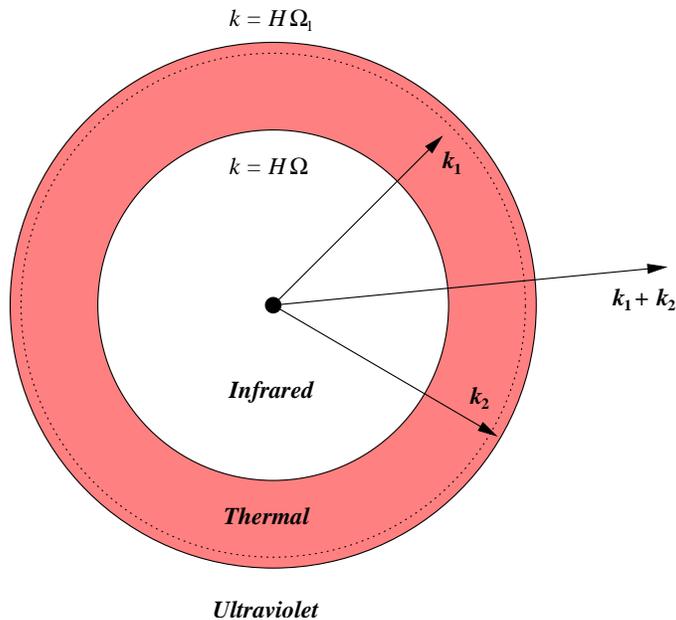,height=3.2in}}
\caption{\footnotesize The 3-sphere of co-moving $k$ modes 
during deflation. The graph is not 
\break \mbox{} \hspace{1.95cm} 
properly scaled.}
\end{figure}

Infrared modes produced during inflation can now scatter 
in the dense bath of thermal modes. Since there are more 
modes as $k$ increases, and, since the thermal modes with 
the highest $k$ appear first, it is more likely for an 
infrared mode to scatter with high $k$ thermal modes. This 
scattering process can increase the wave number 
$k < H \Omega$ of the infrared mode and turn it into a 
wave number $k > H {\it \Omega}_1$ of an ultraviolet mode 
{\it (see Figure 6)}. Hence, the population of infrared 
modes or, equivalently, the amplitude of their mode 
functions decreases. It is this decrease that is responsible 
for terminating the growth of the strong infrared effect 
right after the transition from inflation to deflation occurs. 
Moreover, as we shall see, it ensures that deflation must 
end. \\

{\it (ii) The Infrared Decay Rate} \\
The calculation of the decay rate shall be done in the context 
of the graviton-like scalar model of Section 2. The dominant 
process, to lowest order in perturbation theory, is shown in 
Figure 7. As long as there is a diffuse small density of 
infrared modes -- the production of which is the source of 
the effect -- perturbation theory gives reliable results. 

\begin{figure}
\centerline{\epsfig{file=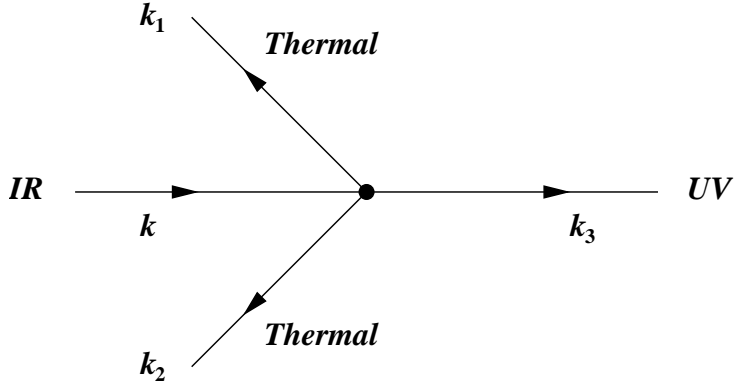,height=2in}}
\caption{\footnotesize The dominant scattering process during 
deflation.}
\end{figure}

In terms of the differential transition probability $dP$, the 
transition rate $\Gamma$ per unit conformal time is given by:
\begin{equation}
\Gamma(\eta, k) = \frac{d}{d\eta} \, \int
dP \left( 
{\vec k} \, \rightarrow \, 
{\vec k}_1 \, , {\vec k}_2 \, , {\vec k}_3 \right) 
\;\; , \label{rate}
\end{equation}
where $\eta$ is the observation conformal time. The differential 
transition probability is obtained from the transition amplitude 
$\mbox{\Large ${\it m}$}$ for the process:
\begin{eqnarray}
dP \left( 
{\vec k} \, \rightarrow \, 
{\vec k}_1 \, , {\vec k}_2 \, , {\vec k}_3 \right)  & = & 
\Vert \, \mbox{\Large ${\it m}$} \left( 
{\vec k} \, \rightarrow \, 
{\vec k}_1 \, , {\vec k}_2 \, , 
{\vec k}_3 \right) \Vert^2 \; \times
\label{diffprob} \\
& \mbox{} &
\frac{d^3{\vec k}_1}{(2\pi)^3} \; 
\frac{d^3{\vec k}_2}{(2\pi)^3} \;  
\frac{d^3{\vec k}_3}{(2\pi)^3} \;
(2\pi)^3 \, 
\delta^3 \left( 
{\vec k} - {\vec k}_1 - {\vec k}_2 - {\vec k}_3 \right) 
\;\; , \nonumber
\end{eqnarray}
where:
\begin{eqnarray}
& \mbox{} &
i \mbox{\Large ${\it m}$} \left( 
{\vec k} \, \rightarrow \, 
{\vec k}_1 \, , {\vec k}_2 \, , {\vec k}_3 \right) \;
(2\pi)^3 \, 
\delta^3 \left( 
{\vec k} - {\vec k}_1 - {\vec k}_2 - {\vec k}_3 \right) \equiv
\label{Mdef} \\
& \mbox{} & \hspace{5.3cm}
\Big\langle 0 \ \Big\vert \;
\alpha_0({\vec k}_1) \;
\alpha_0({\vec k}_2) \;
\alpha_0({\vec k}_3) \; \times 
\nonumber \\
& \mbox{} & \quad
{\bf T} \left\{ \;
\exp \left[ \,
-i \frac{\lambda}{4!} \,
\int_{\eta_1}^{\eta}  d\eta' \; \Omega^4(\eta') \;
\int d^3x' \; \varphi^4(\eta', {\vec x}^{\; \prime}) 
\, \right] \; \right\} \;
\alpha_0^{\dagger}({\vec k}) \;
\Big\vert \ 0 \Big\rangle
\;\; . \nonumber
\end{eqnarray}

Using the expansion (\ref{phi+Phifield}-\ref{thermalphi}) of
the field and the canonical commutation relations (\ref{ccr}),
the transition amplitude takes the following form in terms of 
the mode functions (\ref{u+Umodefncts}):
\begin{eqnarray}
& \mbox{} & 
i \mbox{\Large ${\it m}$} \left( 
{\vec k} \, \rightarrow \, 
{\vec k}_1 \, , {\vec k}_2 \, , {\vec k}_3 \right) =
\label{M} \\
& \mbox{} & \quad
-i \lambda \int_{\eta_1}^{\eta}
d\eta' \; \Omega^4(\eta') \;\;
{\rm u}^*(\eta', k_1) \; {\rm u}^*(\eta', k_2) \;
{\rm u}^*(\eta', k_3) \; {\rm u}(\eta', k)
\; + \; O(\lambda^2)
\;\; . \nonumber
\end{eqnarray}
The form of the mode functions appropriate for the infrared 
wave number $k$, the ultraviolet wave number $k_3$ and the 
thermal wave numbers $k_{1,2}$, was obtained in
(\ref{deflirmodefncts}-\ref{deflthermodefncts}):
\begin{eqnarray}
{\rm u}(\eta, k) & = & 
{\rm u}_{\rm IR}(\eta, k) \; \sim \;
- \frac{i H}{\sqrt {2k^3}} 
\; \; , \label{irmodefnct} \\ 
{\rm u}(\eta, k_3) & = & 
{\rm u}_{\rm UV}(\eta, k_3) \; \sim \;
\frac{\Omega^{-1}}{\sqrt {2k_3}} \;
e^{-ik_3\eta} 
\; \; , \label{uvmodefnct} \\
{\rm u}(\eta, \, k_{1,2}) & = &
{\rm u}_{\rm TH}(\eta, \, k_{1,2}) \; \sim \;
\sqrt \frac2{k_{1,2}} \; \frac{i H}{k_{1,2}} \; 
\frac{{\it \Omega}_1}{\Omega} \; 
{\cos \left( \frac{k_{1,2}}{H \Omega} \right)} 
\; \; . \label{thermalmodefnct} 
\end{eqnarray}
The observation time is taken to be well in-between the onset
of inflation at $-H^{-1}$ and the transition time from inflation
to deflation at $\eta_1$: 
\footnote{Recall the relation (\ref{conftranstimevalue}) between
$\eta_1$ and $H^{-1}$.}
\begin{equation}
\vert \eta_1 \vert \ll \eta \ll H^{-1} 
\; \; . \label{timeseq} 
\end{equation}
Typical wave numbers associated with the modes of the process
satisfy:
\begin{equation}
H \ll k \ll H \Omega \ll k_{1,2} \ll H {\it \Omega}_1 \ll k_3
\; \; . \label{modeseq} 
\end{equation}
By substituting (\ref{irmodefnct}-\ref{thermalmodefnct}) in 
(\ref{M}), the transition rate becomes:
\begin{eqnarray}
\Gamma(\eta, k) \; \sim \;
\lambda^2 \, H^6 \, {\it \Omega}_1^4 \;
\frac{\Omega}{k^3} \;
\int_{\eta_1}^{\eta} d\eta' \; \Omega(\eta') \; 
\int \frac{d^3{\vec k}_1}{(2\pi)^3} \; 
\frac{\cos (k_1 \eta) \, \cos (k_1 \eta')}{k_1^3} \; \times
\nonumber \\
\int \frac{d^3{\vec k}_2}{(2\pi)^3} \; 
\frac{\cos (k_2 \eta) \, \cos (k_2 \eta')}{k_2^3} \;\; 
\frac{\cos \left[ \, \Vert {\vec k}_1 + {\vec k}_2 \Vert \,
(\eta - \eta') \, \right]}{\Vert {\vec k}_1 + {\vec k}_2 \Vert}
\;\; , \quad \label{rate2}
\end{eqnarray}
where we used:
\begin{equation}
{\vec k_3} = {\vec k} - {\vec k_1} - {\vec k_2} 
\; \sim \; - \left( {\vec k_1} + {\vec k_2} \right)
\; \; . \label{k_3sub} 
\end{equation}
Performing the angular integrations and noting that 
oscillatory destructive interference is avoided only 
when $k_{1,2} \ \Delta\eta \leq 1$, gives:
\begin{eqnarray}
\Gamma(\eta, k) \; \sim \;
\frac{\lambda^2 \, H^6 \, {\it \Omega}_1^4}{8\pi^4} \;
\frac{\Omega}{k^3} \;
\int_{t_1}^t dt' \;  
\int_{H \Omega}^{H {\it \Omega}_1} 
\frac{dk_1}{k_1} \; \cos (k_1 \, \Delta\eta) \; \times
\nonumber \\
\int_{H \Omega}^{H {\it \Omega}_1} 
\frac{dk_2}{k_2} \, \cos (k_2 \, \Delta\eta) \;
\frac{\cos (k_1 \, \Delta\eta) \; \sin (k_2 \, \Delta\eta)}
{k_1 \, k_2 \, \Delta\eta}
\;\; . \label{rate3}
\end{eqnarray}
The final answer is:
\begin{equation}
\Gamma(\eta, k) \; \sim \; 
\frac{\lambda^2 \; H^4 \; {\it \Omega}_1^4}
{24\pi^4 \; k^3}
\;\; . \label{rateconf}
\end{equation}
It is elementary to recover the transition rate in co-moving 
time:
\begin{equation}
\Gamma(t, k) \; = \; 
\Omega^{-1} \; \Gamma(\eta, k)
\;\; , \label{ratecomove}
\end{equation}
and the decay rate $\mbox{\Large $\tau$}$:
\begin{equation}
\frac1{\mbox{\Large $\tau$}}(t, k) \; = \;
\frac{\Gamma(t, k)}{N_{\rm IR}({\vec k} \, ; t)} 
\;\; . \label{decayrate}
\end{equation}
Using (\ref{ratecomove}) and (\ref{irmodesnumber}), we obtain:
\begin{equation}
\frac1{\mbox{\Large $\tau$}}(t, k) \; \sim \; 
\frac{\lambda^2 \, H^2}{6\pi^4} \;
\frac{{\it \Omega}_1}{k} \; 
\left( \frac{{\it \Omega}_1}{\Omega} \right)^3 \; = \; 
\frac{\lambda^2 \, H^2}{6\pi^4} \;
\frac{e^{H t_1}}{k} \; e^{3H (t-t_1)}
\;\; , \label{decayrate2}
\end{equation}
and it becomes apparent that the process becomes very strong
very quickly. A way to understand this comes from considering
the volume of the co-moving modes sphere {\it (see Figure 6)}. 
Since the number of modes of wave number $k$ grows dramatically 
with increasing $k$, almost all the volume of the sphere is 
concentrated close to the surface at $k = H {\it \Omega}_1$:
\begin{equation}
{\rm V_{skin}} \ = \
\frac43 \pi \left( H {\it \Omega}_1 \right)^3
\left[ \ 1 - e^{- H {\it \Delta} t} \ \right] \; \sim \;
\frac43 \pi \left( H {\it \Omega}_1 \right)^3
\ = \ {\rm V_{sphere}}
\;\; . \label{Vskin}
\end{equation}

{\it (iii) The Induced Stress} \\
During inflation, the source of the induced stress tensor was
the production of infrared particles. The objective is to determine 
the evolution of the induced stress tensor after the transition 
to deflation. As it was just shown, in the deflationary regime 
there is a strong depletion of the infrared particles due to the 
presence of a thermal particle bath with which they can scatter 
and become ultraviolet. Accordingly, the mode functions during
deflation must be degraded to account for this depletion.

This is properly achieved by solving for the propagator in the
presence of the stochastic background (\ref{thermalphivalue}).
In this case, there are no more plane wave momentum states.
Nonetheless, a reasonable approximation is obtained by noting 
that the dominant effect comes from coherent interactions
between infrared modes and is diminished upon their depletion. 
Therefore, we use plane wave states which then scatter with
the background and disappear from the population of infrared
modes. Each mode function is effectively degraded by the 
exponential of the integral of minus the decay rate:
\begin{equation}
{\bar {\rm u}}(t, k) \; \equiv \;
\exp \left[ - \frac12 \int_{t_1}^t dt' \; 
\frac1{\mbox{\Large $\tau$}}(t', k) \, \right] \; 
{\rm u}(t, k)
\;\; . \label{degradedmodefnct}
\end{equation}
The degradation factor can be estimated using (\ref{decayrate2}):
\begin{equation}
\exp \left[ - \frac12 \int_{t_1}^t dt' \; 
\frac1{\mbox{\Large $\tau$}}(t', k) \, \right] \; \sim \;
- \frac{\lambda^2 \, H}{2^2 \, 3^2 \, \pi^4} \;
\frac{e^{H t_1}}{k} \; e^{3H (t-t_1)}
\;\; . \label{degradefactor}
\end{equation}

\begin{figure}
\centerline{\epsfig{file=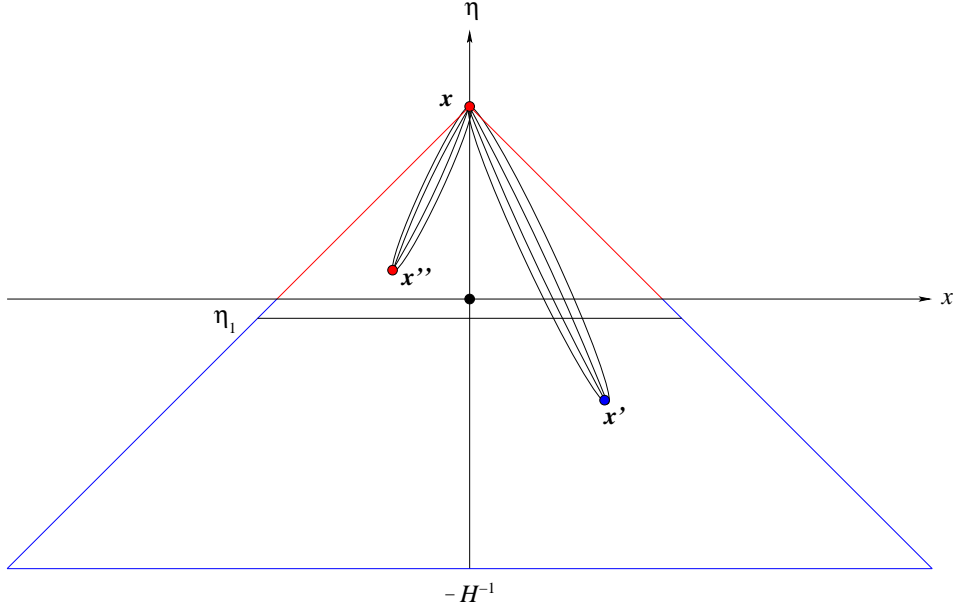,height=3.2in}}
\caption{\footnotesize The dominant diagrammatic contribution to 
the induced ${\rm T}_{\mu\nu}^{\lambda^2}$ during
\break \mbox{} \hspace{2cm} 
deflation. The graph is not properly scaled.}
\end{figure}

Consider the first non-trivial contribution to the induced 
stress tensor. Its leading part:
\begin{equation}
{\bar {\rm T}}_{\mu\nu}^{\, \colon \varphi \colon} \ \sim \
 - g_{\mu\nu} \; \left\langle 0 \left\vert 
\; \frac{1}{4!} \lambda \; \colon \varphi^4 \colon \;
\right\vert 0 \right\rangle  
\;\; , \label{leadingTmn2}
\end{equation}
has the same topology it did during inflation 
{\it (see Figure 8)}. One of the two vertices is at the 
observation event at $x = (\eta, {\vec 0})$. The other 
can be anywhere within the past lightcone of the observer 
and there are two possibilities. In the first, the vertex 
lies in the deflationary part -- for instance, at $x''$ -- 
and the mode functions both at $x$ and $x''$ must be degraded. 
But when the vertex is situated in the inflationary region 
-- for instance, at $x'$ -- only the mode functions at $x$ 
must be degraded. Hence, it is the latter case that dominates 
and we shall concentrate on the corresponding diagram.

The exact calculation of the diagram using the degraded mode 
functions ${\bar {\rm u}}(t, k)$ at $x$, may not be feasible. 
A prohibitive fact is the presence of $k$ dependence in 
(\ref{degradefactor}). It is apparent from (\ref{degradefactor}), 
that the thermal screening effect caused by the presence of 
the degrading factor gets stronger with decreasing $k$. At 
the same time, the largest contribution to the induced stress 
tensor comes from infrared modes whose wave number is close 
to $k \sim H \Omega$. Thus, calculability can be restored by 
evaluating (\ref{degradefactor}) at the wave number with the 
maximum impact: $k = H \Omega$. Such a procedure underestimates 
the thermal screening for $k < H \Omega$ and overestimates it 
for $k > H \Omega$. Nonetheless, these two deviations -- being 
opposite in sign -- can balance against each other. In this 
approximation, the desired answer is obtained from the one using 
${\rm u}(t, k)$ by simple multiplication with an appropriate 
overall degrading factor:
\begin{equation}
{\bar {\rm T}}_{\mu\nu}^{\, \colon \varphi \colon} \; \sim \; 
{\rm T}_{\mu\nu}^{\, \colon \varphi \colon} \;
\exp \left[ - 2 \int_{t_1}^t dt' \; 
\frac1{\mbox{\Large $\tau$}}(t', k) \, \right]_{k = H \Omega}
\;\; . \label{degradedTmn}
\end{equation}
The factor of 2 in the exponent of the degrading factor in 
(\ref{degradedTmn}) reflects the four mode functions present 
at $x$. From (\ref{degradefactor}), it trivially follows that:
\begin{equation}
\exp \left[ - 2 \int_{t_1}^t dt' \; 
\frac1{\mbox{\Large $\tau$}}(t', k) \, \right]_{k = H \Omega} 
\; \sim \;
- \frac{\lambda^2}{9 \, \pi^4} \; e^{4 H (t-t_1)}
\; = \;
- \frac{\lambda^2}{9 \, \pi^4} 
\left( \frac{{\it \Omega_1}}{\Omega} \right)^4
\;\; . \label{apprdegradefactor}
\end{equation}

A straightforward application of the Schwinger-Keldysh  rules
to the dominant diagram of Figure 8, results in the following
expression for the expectation value the diagram represents:
\footnote{ The detailed adaptation of the Schwinger-Keldysh
formalism \cite{inin} to the computation of expectation values 
in a cosmological context can be found in \cite{nctrpw1}.} 
\begin{eqnarray}
& \mbox{} & \hspace{-1.5cm}
{\rm T}_{\mu\nu}^{\, \colon \varphi \colon} \, (x) \; = \;
\Omega^2 \eta_{\mu\nu} \; \;
\frac{i \lambda^2}{4!} \;
\int d^4x' \; {\it \Omega}^4(x') \; \times
\nonumber \\
& \mbox{} & \hspace{3cm}
\left\{ \; \left[ \; i \Delta_{++}(x; x') \, \right]^4 
- \, \left[ \; i \Delta_{+-}(x; x') \, \right]^4 \; \right\}
\;\; . \label{leadingTmndefl}
\end{eqnarray}
Due to their correlated interactions, infrared modes comprise 
the ultimate source for the generation of a strong induced stress 
tensor. Under evolution into the deflationary regime, the mode 
functions of the still infrared modes remain unaltered from their 
inflationary form except for the sign change due to the Hubble 
parameter reflection. Therefore, it is again the logarithmic term 
in the propagator (\ref{dSprop}) that carries the leading effect:
\begin{equation}
i\Delta_{++}(x; x') \; \sim \;
- \frac{H^2}{8\pi^2} \; 
\ln \left[ H^2 \left( 
{\it \Delta} x^2 - ({\it \Delta}\eta - i\epsilon)^2 \right)
\right]
\;\; , \label{++logpart}
\end{equation}
\begin{equation}
i\Delta_{+-}(x; x') \; \sim \;
- \frac{H^2}{8\pi^2} \; 
\ln \left[ H^2 \left( 
{\it \Delta} x^2 - ({\it \Delta}\eta + i\epsilon)^2 \right)
\right]
\;\; , \label{+-logpart}
\end{equation}
by providing, eventually, the maximum number of $\ln\Omega$
factors.

Decomposing into real and imaginary parts provides the leading
contribution from the propagator combination appearing in
(\ref{leadingTmndefl}):
\begin{eqnarray}
& \mbox{} & \hspace{-0.9cm}
\left[ \; i \Delta_{++}(x; x') \, \right]^4 - \,
\left[ \; i \Delta_{+-}(x; x') \, \right]^4 \; \sim \; 
\nonumber \\
& \mbox{} & \hspace{-0.6cm}
\sim \; \frac{H^8}{2^{12} \, \pi^8} \; 
\Big\{ \;
\ln \left[ H^2 \left( 
{\it \Delta}\eta^2 - {\it \Delta} x^2 \right) + i\pi \right]^4
- \ln \left[ H^2 \left( 
{\it \Delta}\eta^2 - {\it \Delta} x^2 \right) - i\pi \right]^4
\Big\}
\nonumber \\
& \mbox{} & \hspace{-0.6cm}
\sim \; \frac{H^8}{2^{12} \, \pi^8} \; 
8 \pi i \, 
\ln^3 \left[ H^2 \left(
{\it \Delta}\eta^2 - {\it \Delta} x^2 \right) \right]
\;\; . \qquad \label{logparts}
\end{eqnarray}
Using (\ref{logparts}) and performing the angular integrations 
present in (\ref{leadingTmndefl}) gives:
\begin{eqnarray}
& \mbox{} & \hspace{-2cm}
{\rm T}_{\mu\nu}^{\, \colon \varphi \colon} \, (\eta) \; \sim \;
- \Omega^2 \eta_{\mu\nu} \; \;
\frac{\lambda^2 \, H^8}{2^{10} \, 3 \, \pi^6} \;
\int_{-\frac1{H}}^{\eta_1} d\eta' \; {\it \Omega}^4(\eta') 
\; \times
\nonumber \\
& \mbox{} & \hspace{3.3cm}
{\it \Delta}\eta^3 \int_0^1 dy \; y^2 \; 
\ln^3 \left[ H^2 {\it \Delta}\eta^2 \left( 1 - y^2 \right) \right]
\;\; , \label{leadingTmndefl2}
\end{eqnarray}
where we have rescaled the radial variable 
$r \in \left[ \, 0 \, , \, {\it \Delta}\eta \, \right]$ thusly:
\begin{equation}
r \equiv ({\it \Delta}\eta) \; y 
\quad , \quad 
y \in \left[ \, 0 \, , \, 1 \, \right]
\;\; . \label{radialrescale}
\end{equation}
To leading order, the $y$ integration is:
\begin{eqnarray}
\int_0^1 dy \; y^2 \; 
\ln^3 \left[ H^2 {\it \Delta\eta}^2 \left( 1 - y^2 \right) \right]
& \sim & 
2^3 \; \ln^3 \left( H {\it \Delta\eta} \right) 
\int_0^1 dy \; y^2 
\nonumber \\
& = &  
\frac{2^3}3 \; \ln^3 \left( H {\it \Delta\eta} \right) 
\;\; , \label{radialint}
\end{eqnarray}
so that we get:
\begin{equation}
{\rm T}_{\mu\nu}^{\, \colon \varphi \colon} \, (\eta) \; \sim \;
- \Omega^2 \eta_{\mu\nu} \; \;
\frac{\lambda^2 \, H^4}{2^7 \, 3^2 \, \pi^6} \;
\int_{-\frac1{H}}^{\eta_1} \; \frac{d\eta'}{\eta'^{\, 4}} \; 
\Delta\eta^3 \;
\ln^3 \left( H {\it \Delta\eta} \right) 
\;\; . \label{leadingTmndefl3}
\end{equation}
Because of (\ref{timeseq}) and the domination of the conformal 
time integration from the region close to its upper limit 
$\eta' \sim \eta_1$, we have: 
\begin{eqnarray}
\int_{-\frac1{H}}^{\eta_1} \; \frac{d\eta'}{\eta'^{\, 4}} \; 
{\it \Delta}\eta^3 \;
\ln^3 \left( H {\it \Delta}\eta \right) 
& \sim & 
\eta^3 \; \ln^3(H \eta) \;
\int_{-\frac1{H}}^{\eta_1} \; \frac{d\eta'}{\eta'^{\, 4}}  
\nonumber \\
& = &  
-\frac13
\left( \frac{\eta}{\eta_1} \right)^3 \; 
\ln^3 (H \eta) 
\;\; . \label{conftimeint}
\end{eqnarray}
Consequently:
\begin{equation}
{\rm T}_{\mu\nu}^{\, \colon \varphi \colon} \, (\eta) \; \sim \;
\Omega^2 \eta_{\mu\nu} \; \;
\frac{\lambda^2 \, H^4}{2^7 \, 3^3 \, \pi^6} \;
\left( \frac{{\it \Omega}_1}{\Omega} \right)^3 \; 
\ln^3 \Omega 
\;\; . \label{leadingTmndefl4}
\end{equation}

The physically relevant induced stress tensor during deflation 
is obtained by substituting (\ref{leadingTmndefl4}) and 
(\ref{apprdegradefactor}) in (\ref{degradedTmn}):
\begin{equation}
{\bar {\rm T}}_{\mu\nu}^{\, \colon \varphi \colon} \, (\eta) \; \sim \;
g_{\mu\nu} \; \;
\frac{\lambda^2 \, H^4}{2^7 \, 3^3 \, \pi^6} 
\; \times \; \left( \frac{{\it \Omega}_1}{\Omega} \right)^3 
\; \times \; \ln^3 \Omega 
\; \times \; e^{- \frac{\lambda^2}{9 \, \pi^4} \; 
\left( \frac{{\it \Omega}_1}{\Omega} \right)^4} 
\;\; , \label{degradedleadingTmn1}
\end{equation}
or, in co-moving coordinates:
\begin{equation}
{\bar {\rm T}}_{\mu\nu}^{\, \colon \varphi \colon} \, (t) \; \sim \;
g_{\mu\nu} \; \;
\frac{\lambda^2 \, H^4}{2^7 \, 3^3 \, \pi^6} 
\; \times \; e^{3 H (t - t_1)} 
\; \times \; (H t_1)^3 \; \times \;
e^{- \frac{\lambda^2}{9 \, \pi^4} \; e^{4 H (t - t_1)}}
\;\; . \label{degradedleadingTmn2}
\end{equation}
The induced energy density ${\varrho}$ and pressure ${\rm p}$ 
present in ({\ref{degradedleadingTmn2}) follow directly from 
({\ref{defrho,p}):
\begin{eqnarray}
{\varrho}^{\colon \varphi \colon} \, (t) & \sim &
- \ \frac{\lambda^2 \, H^4}{2^7 \, 3^3 \, \pi^6} \;\;
(H t_1)^3 \;\; e^{3 H (t - t_1)} \;\; 
e^{- \frac{\lambda^2}{9 \, \pi^4} \; e^{4 H (t - t_1)}}
\;\; , \label{pertrhodefl} \\
{\rm p}^{\colon \varphi \colon} \, (t) & \sim & 
- {\varrho}^{\colon \varphi \colon} \, (t) 
\;\; . \label{pertpdefl} 
\end{eqnarray}
The effective Hubble parameter obeys (\ref{defH}) and equals:
\begin{equation}
{\rm H}^{\colon \varphi \colon}(t) \; \sim \; 
H \; \left\{ \; 1 - 
\frac{\lambda^2 \, \varepsilon}{2^5 \, 3^4 \, \pi^4} \;\; 
(H t_1)^3 \;\; e^{3 H (t - t_1)} \;\;
e^{- \frac{\lambda^2}{9 \, \pi^4} \; e^{4 H (t - t_1)}}
\; \right\} 
\;\; . \label{pertHdefl} 
\end{equation}

{\it (iv) The Deflationary Rule} \\
There are three time dependent terms present in the suggestive
expression (\ref{degradedleadingTmn2}) for the induced stress
tensor. And each one of them carries a physical meaning: \\
-- The first represents the causal volume (\ref{vplclimit})
accessed by the deflationary observer. It is an exponentially
growing term and it reflects the vast increase of the volume 
within which interactions that occured from the onset of 
inflation onwards are becoming ``visible''. \\ 
-- The second term contains the effect of infrared particle
production on the observer. This is an inflationary era 
phenomenon and, therefore, its power law growth 
(\ref{leadingTmnresult2}) occurs during that regime and 
terminates at $t_1$. Throughout the inflationary period, 
infrared particle production resulted in a stronger effect 
than the available causal volume; see (\ref{leadingTmnresult2}). 
During deflation, the reverse is true and by a wider margin. \\
-- The third term represents the dense barrier erected by
the depletion of infrared modes as they scatter in the
thermal particle bath created during deflation. As a result,
correlations reaching the observer from the inflationary
epoch are weakened. It overwhelms the other two terms and
weakens the induced stress tensor. 

In analogy to the inflationary conclusion (\ref{inflTmninwords}), 
the induced stress tensor (\ref{degradedleadingTmn2}) can be 
thought of as the combined result of the same geometrical effect 
-- the causal volume accessed by the observer -- and a different 
dynamical effect -- the thermal screening of correlations:
\begin{eqnarray}
{\it Deflationary \; Induced \; Stress} \; \sim 
\hspace{6.7cm} \label{deflTmninwords} \\
{\it ``Causal \; Volume"} \; \times \;
{\it ``Correlated \; Interactions \; Thermal \; Screening"} 
\;\; . \nonumber 
\end{eqnarray}

{\it (v) The End of Deflation} \\
Unlike inflation, during which both dominant effects are 
growing and superpose, in (\ref{deflTmninwords}) they oppose 
each other. Hence, it is possible for their conflicting 
tendencies to cancel out and for balance to be achieved. 
For instance, this happens -- albeit in a trivial sense -- 
at the time $t_1$ of the transition when they are both 
negligible and the induced stress tensor is the one 
developed during the inflationary era.

If there is a later time $t_2 > t_1$ at which balance is 
established again, it will signal the end of deflation. 
The existence of such a time instant is guaranteed by the 
relative strength of the two competing effects. As soon as 
deflation is established, the past lightcone volume of the 
observer grows exponentially and provides the arena for the 
creation of the thermal barrier. Moreover, since thermal
modes redshift like radiation, their density increases as
the inverse fourth power of the scale factor and, therefore,
the scattering source behind the erection of the thermal
barrier increases without bound at a tremendous rate. 
Because the causal volume strengthens the induced stress 
tensor at a slower rate than that with which the barrier 
weakens it, balance between the two effects will unavoidably 
be restored again.

\begin{figure}
\centerline{\epsfig{file=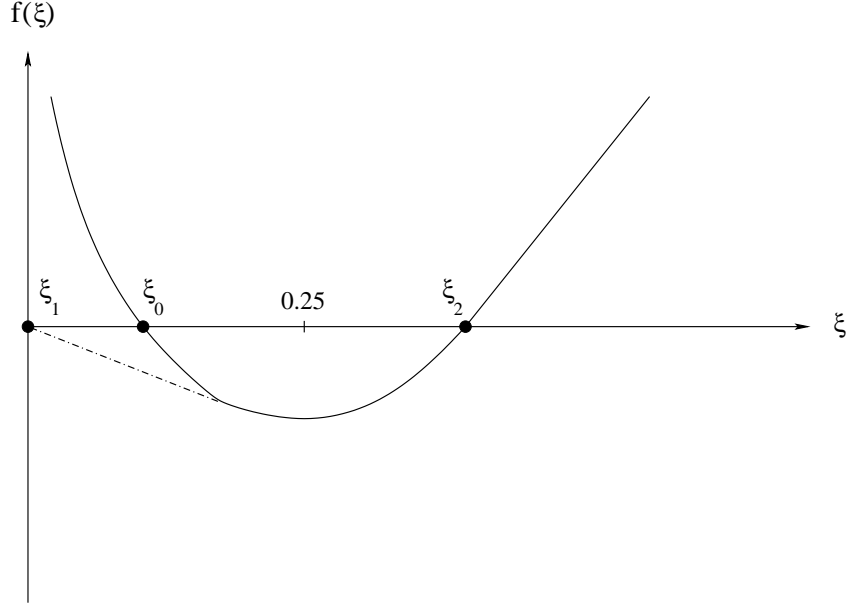,height=3.2in}}
\caption{\footnotesize The function $f(\xi)$ and a possible 
modification {\it (fragmented segment)}
\break \mbox{} \hspace{2.3cm} 
in the region close to $\xi = 0$. The graph is not properly 
scaled.}
\end{figure}

Although the argument showing that deflation has to stop 
follows for quite general reasons which must certainly 
persist in a more exact approach, it is very useful to 
do the analysis within the framework of our present 
approximations. Consider the equation enforcing the balance 
under investigation, as it emerges from 
(\ref{degradedleadingTmn2}): 
\begin{equation}
e^{3 H (t - t_1)} \; \times \;
e^{- \frac{\lambda^2}{9 \, \pi^4} \; e^{4 H (t - t_1)}}
\sim 1
\quad \Longrightarrow \quad
3 H \, (t - t_1) \; \sim \;
\frac{\lambda^2}{9 \pi^4} \; e^{4 H (t - t_1)}
\;\; , \label{deflend1}
\end{equation}
and change to a dimensionless variable:
\begin{equation}
\xi \equiv H \, (t - t_1)
\quad \Longrightarrow \quad
f(\xi) \, \equiv \,
\xi \, - \, \frac14 \, \ln (3 \xi) \, - \,
\frac14 \, \ln \left( \frac{9 \pi^4}{\lambda^2} \right) 
\, = \, 0
\;\; . \label{fdef}
\end{equation}
The function $f(\xi)$ has a minimum at which we require it
to be negative:
\begin{equation}
\xi_{\rm min} \, = \, \frac14
\qquad , \qquad
f \left( \xi_{\rm min} \right) \, <  \, 0
\quad \Longrightarrow \quad
4 e \, < \, \frac{3 \pi^4}{\lambda^2} 
\;\; . \label{fmin}
\end{equation}
The resulting condition is satisfied for all natural values 
of $\lambda$ and there are two non-trivial balancing solutions
{\it (see Figure 9)}. Of these, $\xi_2$ is the desired physical 
solution corresponding to the end of deflation. The function 
$f(\xi)$ cannot capture the other physical solution it should 
have at $\xi_1 = 0$ corresponding to the beginning of deflation; 
near the transition it is untrustworthy and receives corrections. 
The apparent solution at $\xi_0$ is an artifact of the assumption 
we made of being well into the deflationary epoch when computing 
the scattering rate. The solution at $\xi_2$ is clearly within 
the range of validity of our assumption:
\begin{equation}
{\xi}_2 \; \sim \;
\frac14 \; \ln \left( \frac{9 \pi^4}{\lambda^2} \right)
\;\; . \label{xi2}
\end{equation}
Since $t_2$ represents the end of deflation, $\xi_2$ is the
number of deflationary e-foldings ${\rm N}_2$:
\begin{equation}
{\it Deflation \; End} 
\quad \Longrightarrow \quad
{\rm N}_2 \equiv
H \, (t_2 - t_1) \; \sim \;
\frac14 \; 
\ln \left( \frac{9 \pi^4}{\lambda^2} \right)
\;\; . \label{deflduration}
\end{equation}
For any natural value of $\lambda$, (\ref{deflduration}) gives
a small number of e-foldings. Deflation lasts for a very short 
time. Comparison with the duration of inflation (\ref{Ncr})
gives:
\begin{equation}
N_1 \; \gg \; {\rm N}_2 
\;\; . \label{N1vsN2}
\end{equation}

\begin{figure}
\centerline{\epsfig{file=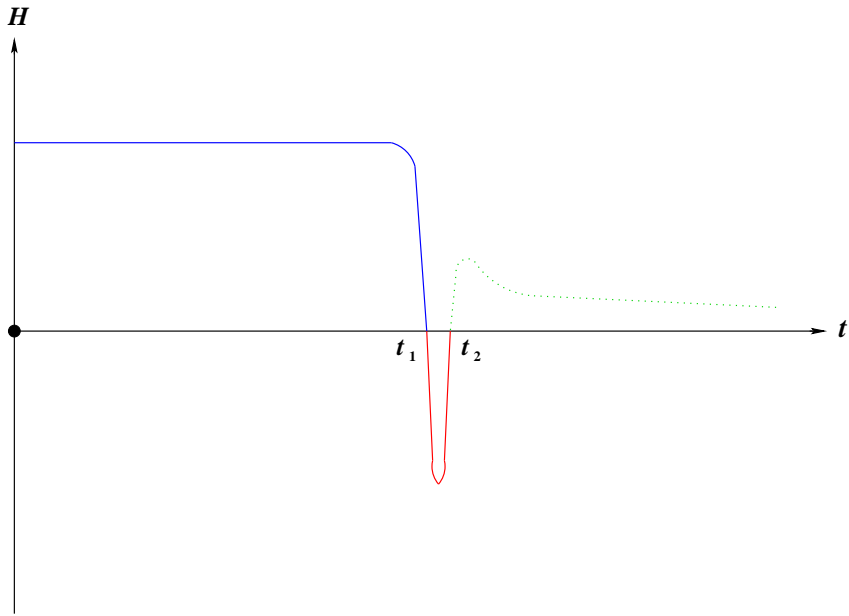,height=3.2in}}
\caption{\footnotesize The expansion rate $H(t)$ during inflation 
{\it (blue)}, deflation {\it (red)}, and 
\break \mbox{} \hspace{2.2cm} 
subsequent expansion {\it (green)}. The graph is not properly scaled.}
\end{figure}

Besides allowing for equations of state with $w < -1$, a brief
period of deflation furnishes a most natural mechanism for the 
reheating of the universe after a long period of inflation.

The kind of post-deflationary regime that emerges needs careful
analysis. Although a succession of inflation to deflation cycles
cannot be excluded, the asymptotic geometry must satisfy:
\begin{equation}
t \gg t_2
\quad \Longrightarrow \quad
H(t) = \ {\dot b} \, (t) \ > \ 0
\;\; . \label{asymptoticH}
\end{equation}
This is a direct consequence of the superior stength of the thermal 
screening effect relative to the causal volume increase. A small
power law expansion, like radiation, is an acceptable post-deflationary
geometry. In Figure 10, where the time evolution of $H(t)$ during
inflation and deflation is exhibited, such a geometry has been 
assumed afterwards and is indicated by the dotted line. \\

\section{Epilogue}

A very important and desirable property of the cosmological evolution
described herein, is its naturalness. $\Lambda$-driven inflation converts
a problem -- the presence of a huge bare cosmological constant in the 
effective theory -- into a virtue. It completely avoids the problem that
starting scalar-driven inflation requires a vanishingly improbable
homogeneous fluctuation over more than a horizon volume \cite{vach1}.

The quantum gravitational back-reaction to inflationary particle
production offers an attractive mechanism for stopping $\Lambda$-driven
inflation. It obviates the need for fine tuning by making the long
duration of inflation a trivial consequence of the fact that gravity
is a weak interaction, even at scales $M \sim 10^{16} GeV$. This same 
fact serves to explain why the induced pressure must be nearly minus 
the induced energy density. The sign of the effect follows from the 
fact that gravity is attractive.
    
Back-reaction certainly slows inflation by an amount which eventually 
becomes non-perturbatively large \cite{nctrpw1}. The problem has long 
been evolving this highly attractive model beyond the breakdown of 
perturbation theory to the current epoch. In this paper, we have 
developed a technique for accomplishing this. The first key idea 
is that, even when the net screening effect becomes large, its 
instantaneous rate of increment is always small. Therefore, it 
can be computed perturbatively. The second key idea is to follow 
Starobinsky \cite{starob1} in subsuming the non-perturbative 
aspects of back-reaction into a stochastic background whose 
evolution obeys the classical field equations.

The screening effect at any point derives from the coherent
superposition of interactions from within the past lightcone 
of that point. Since the invariant volume of the past lightcone
actually grows faster as the expansion rate slows, screening must
overcompensate the bare cosmological constant and lead to a 
period of deflation. Were the physics completely classical and 
stable, this period of deflation would never end. What allows 
the universe to recover is the formation of a thermal barrier of 
the most recently created particles. This barrier scatters the 
virtual infrared quanta needed to propagate the screening effect 
out from the period of inflationary particle production. Because 
the thermal barrier becomes hotter and denser as deflation progresses,
this scattering process must very quickly arrest deflation and allow
slower expansion to resume. That, in turn, dilutes the thermal
barrier, and the subsequent evolution is controlled by the balance
between the growth of the visible portion of the period of
inflationary particle production and the persistence of the thermal
barrier. It is clear that dynamical stability against uncontrollable 
expansion or contraction has been achieved {\it (see Figure 11)}. 
 
\begin{figure}
\centerline{\epsfig{file=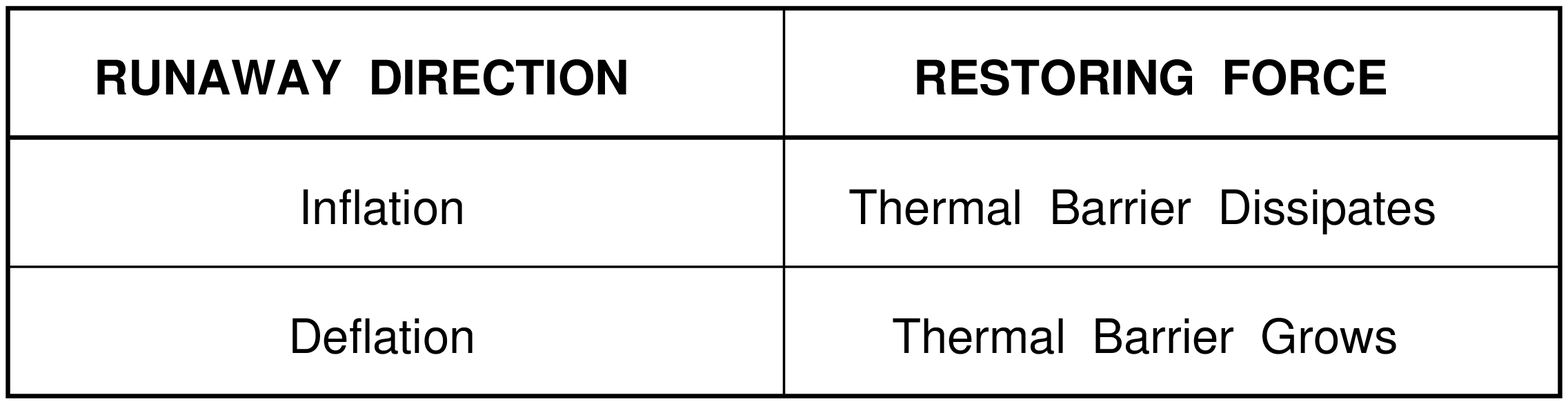,height=1.4in}}
\caption{\footnotesize The physical effects that prevent instabilities 
under time evolution.}
\end{figure}

As far as the construction of a cosmological model is concerned, 
there are some issues that should be addressed: \\
-- During inflation, the model presented is realistic since it was 
entirely based on the graviton. In the post-inflationary analysis,
the graviton-like scalar was used to facilitate the explicit
computations. Nonetheless, the main conclusion -- the existence 
of a short period of deflation -- is universal. The same is true
for the physical role of the various modes: the infrared modes 
carry correlations from the inflationary regime to later times 
and provide the dominant contribution to the quantum induced stress
tensor, the thermal modes decrease the number of infrared modes 
present by scattering with them, and the ultraviolet modes appear 
only as virtual particles. What will most certainly differ is the 
explicit form of the infrared decay rate and induced stress. For a 
realistic cosmology, these computations should be repeated using 
the effective gravitational theory (\ref{Lgr}) and the resulting 
post-deflationary geometry should be determined. Within the same 
context, since the era of deflation furnishes a very natural 
mechanism for reheating, the corresponding temperature should be 
evaluated. \\
-- The dynamical principle which determines evolution is the 
balance between the two restoring forces of Figure 11. As long 
as the expansion -- or contraction -- of spacetime is very rapid, 
gravitation dominates. Matter becomes important when the expansion 
rate does not dominate over a typical matter reaction time. The
desired post-deflationary regime is that of radiation. The 
transition to a matter dominated universe follows naturally. 
Between every pair of successive transitions, there is a period 
of varying imbalance or, equivalently, of varying screening of 
$\Lambda$. If the recently observed data supporting an accelerated
expansion \cite{supernova} persists, it could be the result of
a corresponding imbalance in the dynamical stability of the system 
a while after the transition to matter domination. Whether such 
behaviour occurs, can be determined from the time evolution of 
the expansion rate and the deceleration parameter. Because our
parameter space is very small, it should be possible to obtain 
predictive expressions for these quantities. \\
-- During the short period of deflation, violations of the weak
energy condition are allowed and ``exotic'' equations of state
satisfying $w < -1$ become feasible. It is worthwhile to investigate
whether such a phenomenon could leave its signature in specific
observational measurements \cite{eqnstate}. \\
-- A requirement on the model is its consistency with the measured
spectrum of the cosmic microwave background radiation \cite{cobe}. 
Although a detailed analysis of the evolution of the primordial 
density perturbations within the proposed framework is missing, it 
seems that -- if the inflation to deflation transition is sudden 
and the duration of contraction is small -- the spectrum is not 
altered in any significant way \cite{mrb1}. \\

Finally, the ``phase space'' of our basic assumptions should be 
reviewed: \\
-- {\it Perturbation theory validity:} Perturbative corrections
do become strong but not because each elementary interaction gets 
to be strong -- as is the case, for instance, in QCD. Here, each 
elementary interaction is {\it always} weak. Corrections become 
strong by the coherent superposition of inherently weak processes 
as their number vastly increases with time. Perturbation theory 
breaks in a very soft way because when the dominant term gets to
be strong, all higher order terms are still weak \cite{nctrpw6}. 
Except for a small interval around the transition, perturbation
theory is valid throughout and its leading part provides the main 
physical effect. \\
-- {\it De Sitter spacetime validity:} If we assume a generic power 
law $a(t) = t^s$, there is a lower bound on the power $s$ coming from 
measurements of the spectral index $n = 1 - 2 s^{-1}$ \cite{power}. 
The bound is $s > 20$ and the exponential is a superb approximation 
to such a high power law. \\
-- {\it Spacetime dimensionality:} Modulo exceptional cases, an
extra dimension is equivalent -- from a four-dimensional perspective
-- to a particle spectrum. A general spectrum was studied and the
graviton emerged as the unique known particle that participates
dominantly to the effects. A scalar could contribute as well if
its mass is substantially lower than the natural scale and if its 
dynamics is not described by (\ref{Lmms}). Furthermore, the 
parameter space of such a scalar must be restrictive enough to 
allow predictions; otherwise, the (higher-dimensional) theory
it resides in cannot be fundamental. \\
-- {\it Supersymmetry:} All the analysis is completely independent 
of supersymmetry. It is, therefore, valid for any supersymmetric
theory both in its unbroken and -- the physically unavoidable --
broken regime.

\newpage

\centerline{\bf Acknowledgements}
 
This work was partially supported by European Union grants 
HPRN-CT-2000-00122 and HPRN-CT-2000-00131, by the DOE contract 
DE-FG02-97ER41029, and by the Institute for Fundamental Theory 
at the University of Florida.

\end{document}